\documentclass[12pt]{article}
\usepackage{a4}
\usepackage[utf8x]{inputenc}
\usepackage{amsmath}
\usepackage{amssymb}
\usepackage{cases}
\usepackage{theorem}
\usepackage{subfigure}
\usepackage{boxedminipage}
\usepackage{graphicx}
\usepackage{float}
\usepackage{rotating}
\usepackage{color}
\usepackage{hyperref}
\usepackage{changebar}
\nochangebars

\newcommand{\card}[1]{\arrowvert #1\arrowvert}

\floatstyle{boxed}
\newfloat{encadre}{htbp}{enc}
\floatname{encadre}{Encadré}

\title{Revisiting old combinatorial beasts in the quantum age: quantum annealing versus maximal matching}
\author{Daniel Vert, Renaud Sirdey and St\'ephane Louise\thanks{\texttt{daniel.vert2@cea.fr}, \texttt{renaud.sirdey@cea.fr}, \texttt{stephane.louise@cea.fr}}
\\ CEA, LIST, France}

\begin{document}

\maketitle

\begin{abstract}
This paper experimentally investigates the behavior of analog quantum computers such as commercialized by D-Wave when confronted to instances of the maximum cardinality matching problem specifically designed to be hard to solve by means of simulated annealing. We benchmark a D-Wave ``Washington'' (2X) with 1098 operational qubits on various sizes of such instances and observe that for all but the most trivially small of these it fails to obtain an optimal solution. Thus, our results suggests that quantum annealing, at least as implemented in a D-Wave device, falls in the same pitfalls as simulated annealing and therefore suggest that there exist polynomial-time problems that such a machine cannot solve efficiently to optimality.
\end{abstract}

\section{Introduction}

From a practical view, the emergence of quantum computers able to compete with the performance of the most powerful conventional computers remains highly speculative in the foreseeable future. Indeed, although quantum computing devices are scaling up to the point of achieving the so-called milestone of quantum supremacy \cite{google-2019}, these intermediate scale devices, referred to as NISQ \cite{preskill-2018}, will not be able to run mainstream quantum algorithms such as Grover, Shor and their many variants at practically significant scales. Yet there are other breeds of machines in the quantum computing landscape, in particular the so-called analog quantum computers of which the machines presently sold by the Canadian company D-Wave are the first concrete realizations. These machines implement a noisy version of the Quantum Adiabatic Algorithm introduced by Farhi et al. in 2001 \cite{farhi-2001}. From an abstract point of view, such a machine may be seen as an oracle specialized in the resolution of an NP-hard optimization problem\footnote{Strictly speaking, to the best of the authors' knowledge, although the general problem is $NP$-hard, the complexity status of the more specialized instances constrained by the qubit interconnection topology of these machines remains open.} (of the spin-glass type) with an algorithm functionally analogous to the well-known simulated annealing but with a quantum speedup (the precise characterization of which still being an open question).

On top of the formal analogies between simulated and quantum annealing, there also appears to be an analogy between the latter present state of art and that of simulated annealing when it was first introduced. So it might be useful to recall a few facts on SA. Indeed, simulated annealing was introduced in the mid-80's~\cite{kirkpatrick-1983,cerny-1985} and its countless practical successes quickly established it as a mainstream method for approximately solving computationally-hard combinatorial optimization problems. Thus, the theoretical computer science community investigated in great depth its convergence properties in an attempt to understand the worst-case behavior of the method. With that respect, these pieces of work, which were performed in the late 80's and early 90's, lead to the following insights. First, when it comes to solving combinatorial optimization problems to optimality, it is necessary (and sufficient) to use a logarithmic cooling schedule~\cite{geman-1984, hajek-1988,nolte-1996} leading to an exponential-time convergence in the worst-case (an unsurprising fact since it is known that $P\neq NP$ in the oracle setting~\cite{baker-1975}). Second, particular instances of combinatorial problems have been designed to specifically require an exponential number of iterations to reach an optimal solution for example on the (NP-hard) 3-coloring problem~\cite{nolte-1996} and, more importantly for this paper, on the (polynomial) maximum cardinality matching problem~\cite{sasaki-1988}. Lastly, another line of works, still active today, investigated the asymptotic behavior of hard combinatorial problems~\cite{burkard-1985,frenck-1985,schauer-2016} showing that the cost ratio between best and worst-cost solutions to random instances tends (quite quickly) to 1 as the instance size tend to $\infty$. These latter results provided clues as to why simple heuristics such as simulated annealing appear to work quite well on large instances as well as to why branch-and-bound type exact resolution methods tend to suffer from a trailing effect (i.e. find optimal or near-optimal solutions relatively quickly but fail to prove their optimality in reasonable time).

Despite these results now being quite well established, they can also contribute to the ongoing effort to better understand and benchmark quantum adiabatic algorithms~\cite{farhi-2001} and especially the machines that now implements it in order to determine whether or not they provide a quantum advantage over their classical counterparts. Still, as it is considered unlikely that any presently known quantum computing paradigm will lead to efficient algorithms for solving $NP$-hard problems, determining whether or not quantum adiabatic computing yields an advantage over classical computing is most likely an ill-posed question given present knowledge. Yet, as a quantum analogue of simulated annealing, attempting to demonstrate a quantum advantage of adiabatic algorithms over simulated annealing appears to be a better-posed question. At the time of writing, this problem is the focus of a lot of works which, despite claims of exponential speedups in specific cases \cite{farhi-2002} (which also lead to the development of the promising Simulated Quantum Annealing classical metaheuristic \cite{crosson-2016}), hint towards a logarithmic decay requirement of the temperature-analog of QA but with smaller constants involved \cite{santoro-2002} leading to only an $O(1)$ advantage of QA over SA in the general case. Such an advantage has furthermore recently been experimentally demonstrated by Albash and Lidar \cite{albash-2018}.
The present paper contributes to the study of the QA vs SA issue by experimentally confronting a D-Wave quantum annealer to the pathological instances of the maximum cardinality matching problem proposed by Sasaki and Hajek~\cite{sasaki-1988} in order to show that simulated annealing was indeed unable to solve certain polynomial problems in polynomial time. Demonstrating an ability to solve these instances to optimality on a quantum annealer would certainly hint towards a worst-case quantum annealing advantage over simulated annealing whereas failure to do so would tend to demonstrate that quantum annealing remains subject to the same pitfalls as simulated annealing and is therefore unable to solve certain polynomial problems efficiently.

As a first step towards this, the present paper experimentally benchmarks a D-Wave ``Washington'' (2X) with 1098 operational qubits on various sizes of such pathologic instances of the maximum cardinality matching problem and observes that for all but the most trivially small of these it fails to obtain an optimal solution. This thus provides negative evidences towards the existence of a worst-case advantage of quantum annealing over classical annealing. As a by-product, our study also provides feedback on using a D-Wave annealer in particular with respect to the size of problems that can be mapped on such a device. This paper is organized as follows. Sect. \ref{sec:qa} provides some background on quantum annealing, the D-Wave devices and their alleged limitations. Sect.~\ref{sec:match:parapluie} surveys the maximum cardinality matching problem, introduces the $G_n$ graph family underlying our pathologic instances and subsequently details how we build the QUBO instances to be mapped on the D-Wave from those instances. Then, Sect.~\ref{sec:exp} extensively details our experimental setup and experimentations and Sect.~\ref{sec:discuss}  concludes the paper with a discussion of the results and a number of perspectives to follow up on this work.


\section{Quantum annealing and its D-Wave implementation}
\label{sec:qa}


\subsection{The generalized Ising problem and QUBO}

D-Wave systems are based on a quantum annealing process\footnote{A combinatorial optimization technique functionally similar to conventional (simulated) annealing but which, instead of applying thermal fluctuations, uses quantum phenomena to search the solution space more efficiently \cite{farhi2000quantum}.} which goal is to minimize the Ising Hamiltonian:

\begin{equation}
    \mathcal{H}(\mathbf{h},\mathbf{J},\boldsymbol\sigma)=\sum_{i}h_{i}\sigma_{i}+\sum_{i<j}J_{ij}\sigma_{i}\sigma_{j},
    \label{eq:objective}
\end{equation}

where the external field $\mathbf{h}$ and spin coupling interactions matrix $\mathbf{J}$ are given, and the vector of spin (or qubit) values $\boldsymbol\sigma/\forall i, \sigma_i\in\{-1, 1\}$ is the variable for which the energy of the system is minimized as the process of adiabatic annealing transition the system from a constant coupling with a superposition of spins\footnote{The initial Hamiltonian is proportional to $\sum_{i,j}\sigma^x_i\sigma^x_j$, hence based on Eigen-vectors of operator $\widehat{\sigma^x}$ (on the $x$-axis) whilst the momentum of spin on Eq.~\ref{eq:objective} is an Eigen-state of $\widehat{\sigma^z}$ (on the $z$-axis) for which Eigen-states of $\widehat{\sigma^x}$ are superposition states. The adiabatic theorem allows transitioning from the initial ferromagnetic state on axis $x$ to an eigen-state of the Hamiltonian of Eq.~\ref{eq:objective} on axis $z$ and hopefully to the lowest energy of it.} to the final Hamiltonian as given by Eq.~\ref{eq:objective}. Historically speaking, the Ising Hamiltonian corresponds to the case where only the closest neighbouring spins are allowed to interact (\emph{i.e.} $J_{ij}\neq 0 \iff$ nodes $i$ and $j$ are conterminous). The generalized Ising problem, for which any pair of spins in the system are allowed to interact, is easily transformed into a well known optimization problem called QUBO (for Quadratic Unconstrained Binary Optimization) which objective function is given by:

\begin{equation}
    O(\text{\textbf{Q}},\text{\textbf{x}})=\sum_{i}Q_{ii}x_i+\sum_{i<j}Q_{ij}x_i x_j,
    \label{eq:qubo}
\end{equation}

in which the matrix $\boldsymbol{Q}$ is constant and the goal of the optimization is to find the vector of binary variables $\forall i, x_i\in\{0,1\}$ that either minimizes or maximizes the objective function $O(\boldsymbol{Q},\boldsymbol{x})$ from Eq.~\ref{eq:qubo}. For the minimization problem (but only a change of sign away for the maximization problem), it is trivial that the generalized Ising problem and the QUBO problem are equivalent given $\forall i, Q_{ii}=h_i$, $\forall i,j/i\neq j, Q_{ij}=J_{ij}$ and $\forall i, \sigma_i= 2 x_i -1$.

Hence, if quantum annealing can reach a configuration of minimum energy, then the associated state vector solves the equivalent QUBO problem at the same time. 
As the behavior of each qubit in a quantum annealer allows them to be in a superposition state (a combination of the states ``$-1$'' and ``$+1$'') until they relax to either one of these eigen-states, it is thought that quantum mechanical phenomena -- e.g., quantum tunneling -- can help reaching the minimum energy configuration, or at least a close approximation of it, in more cases than with Simulated Annealing (SA). 
Indeed, when SA only relies on (simulated) temperatures to pass over barriers of potential, in Quantum Annealing, quantum phenomena can help because tunneling is more efficient to pass energy barriers even in the case where the temperature is low. Therefore, this technique is a promising heuristic approach to ``quickly'' find acceptable solutions for certain classes of complex NP-Hard problems that are easily mapped to these machines, such as optimization, machine learning, or operational research problems.


\subsection{D-Wave limitations}

\begin{figure}[htbp]
\begin{center}
\includegraphics[width=\textwidth]{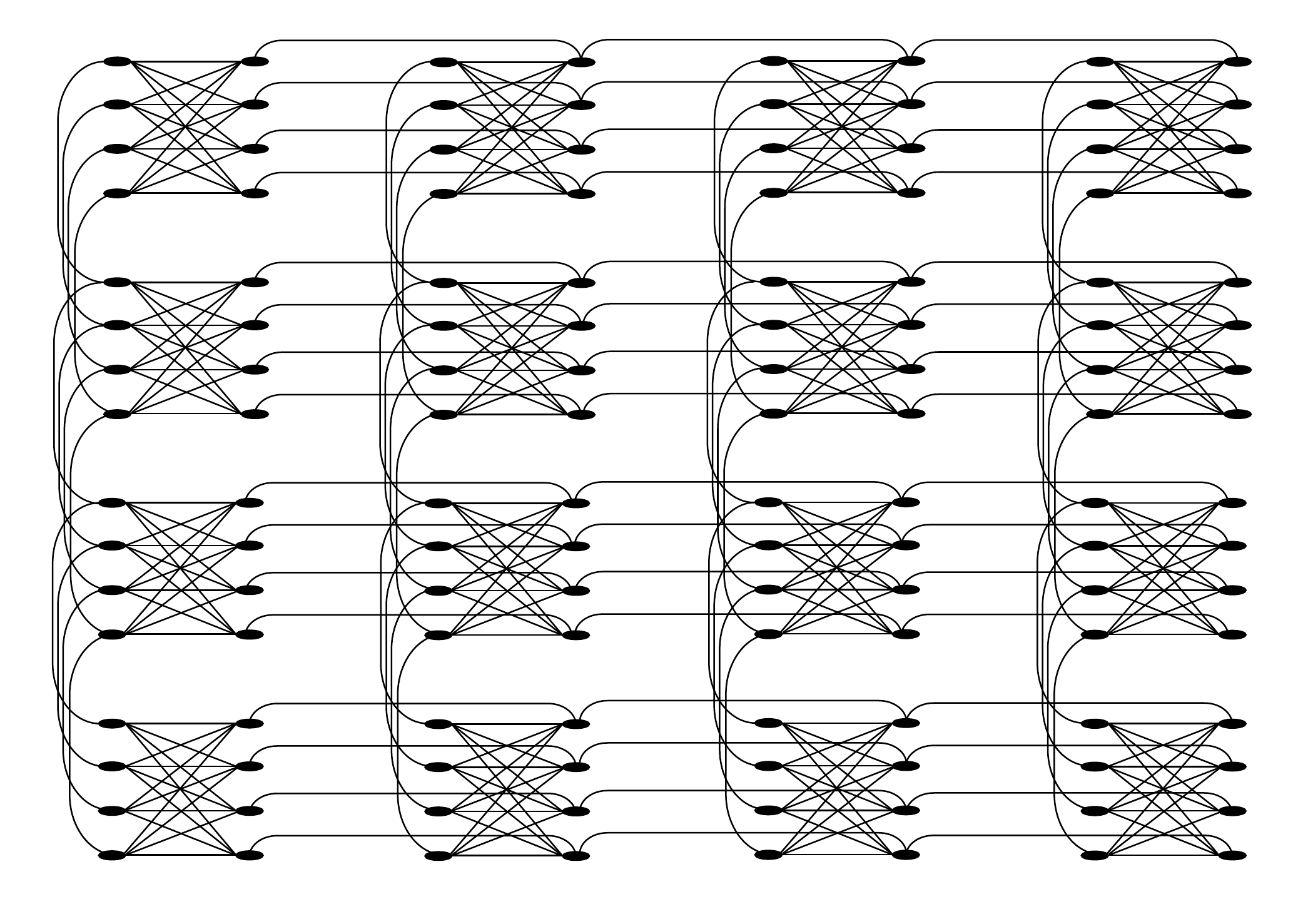}
\end{center}
\caption{Representation of a Chimera graph with $4 \times 4$ unit cells,
each a small $2\times4$ bipartite graph, for $128$ physicals qubits. The links represents all the inter-spin coupling $J_{ij}$ that can be different from 0.}
\label{fig:chimera}
\end{figure}

Nonetheless, it is worth noting, that in the case of the current architectures of the D-Wave annealing devices, the freedom to choose the $J_{ij}$ coupling constants is severely restrained by the hardware qubit interconnection topology.
In particular, this so-called \emph{Chimera} topology is sparse, with a maximum number of inter-spin couplings limited to a maximum of 6 per qubit (or spin variable).
Fig.~\ref{fig:chimera} illustrates an instance of the Chimera graph for $128$ qubits, $T = (N_T, E_T)$, where nodes $N_T$ are qubits and represent problem variables with programmable weights ($h_i$), and edges $E_T$ are associated to the couplings $J_{ij}$ between qubits ($J_{ij}\neq 0 \implies (i,j)\in E_T$). 
As such, if the graph induced by the nonzero couplings is not isomorphic to the Chimera graph, which is the case most usually, then one must resort to several palliatives among which the duplication of logical qubits onto several physical qubits is the least disruptive one if the corresponding expanded problem can still fit on the target device.

Then, a D-Wave annealer minimizes the energy from the Hamiltonian of Eq.~(\ref{eq:objective}) by associating weights ($h_i$) with qubit spins ($\sigma_i$) and couplings ($J_{ij}$) with couplers between the spins of the two connected qubits ($\sigma_i$ and $\sigma_j$).
As an example, the D-Wave 2X system we used has 1098 operational qubits and 3049 operational couplers. 



As said previously, a number of constraints have an impact on the practical efficiency of this type of machines. In~\cite{bian2016mapping}, the authors highlight four factors: the precision/control error which is limited by the parameters $\mathbf{h}$ and $\mathbf{J}$ which value ranges are also limited\footnote{The range of $h_i \in [-2,+2]$ and $J_{i,j} \in [-1,+1]$ is a limitation for all values of the variables to be included in the graph. If the values of $h_i$ and $J_{i,j}$ are outside their respective ranges, then they are unavailable and not mapped}, the low connectivity\footnote{If the problems to be solved do not match the structure of the $T$ graph architecture, then they cannot be mapped and resolved directly.} in $T$, and the in fine small number of useful qubits once the topological constraints are accounted for. In~\cite{bian2014discrete}, the authors show that using large energy gaps in the Ising representation of the model one wants to optimize can greatly mitigate some of the intrinsic limitations of the hardware like precisions over the coupling and noises in the spin measurements. They also suggest using ferromagnetic Ising coupling between qubits (i.e., making qubit duplication) to mitigate the issues with the limited connectivity of the Chimera graph. All these suggestions can be considered good practices (which we did our best to follow) when trying to use the D-Wave machine to solve real Ising or QUBO problems with higher probabilities of outputting the best solution despite hardware and architecture limitations.

Thus, preprocessing algorithms are required to adapt the graph of a problem to the hardware. Pure quantum approaches are limited by the number of variables (duplication included) that can be mapped on the hardware. Larger graphs require the development of hybrid approaches (both classical and quantum) or the reformulation of the problem to adapt to the architecture. For example, for a $128 \times 128$ matrix size, the number of possible coefficients $J_{ij}$ is $8128$ in the worst-case, while the Chimera graph which associates $128$ qubits ($4 \times 4$ unit cells) has ``only'' $318$ couplers. The topology therefore accounts only for $\sim 4\%$ of the total number of couplings required to map a $128 \times 128$ matrix in the worst case. Although preliminary studies (e.g., \cite{vert2019limitations}) have shown that it is possible to obtain solutions close to known minimums for $\mathbf{Q}$ matrices with densities higher than those permitted by the Chimera graph by eliminating some coefficients, they have also shown that doing so isomorphically to the Chimera topology is difficult. It follows that solving large and dense QUBO instances requires nontrivial pre and postprocessing as well as a possibly large number of invocations of the quantum annealer. 


\section{Solving maximum cardinalty matching on a quantum annealer}
\label{sec:match:parapluie}

\subsection{Maximum cardinality matching and the $\mathbf{G_n}$ graph family}
\label{sec:matching}


Given an (undirectered) graph $G=(V,E)$, the maximum matching problem asks for $M\subseteq E$ such that $\forall e,e'\in M^2$, $e\neq e'$ we have that $e\cap e'=\emptyset$ and such that $|M|$ is maximum. The maximum matching problem is a well-known polynomial problem dealt with in almost every textbook on combinatorial optimization (e.g.,~\cite{korte-2012}), yet the algorithm for solving it in general graphs, Edmond's algorithm, is a nontrivial masterpiece of algorithmics. Additionally, when $G$ is bipartite i.e. when there exists two collectively exhaustive and mutually exclusive subsets of $E$, $A$ and $B$, such that no edge has both its vertices in $A$ or in $B$, the problem becomes a special case of the maximum flow problem and can be dealt with several simpler algorithms~\cite{korte-2012}.

\begin{figure}[htbp]
\begin{center}
\includegraphics[width=\textwidth,keepaspectratio]{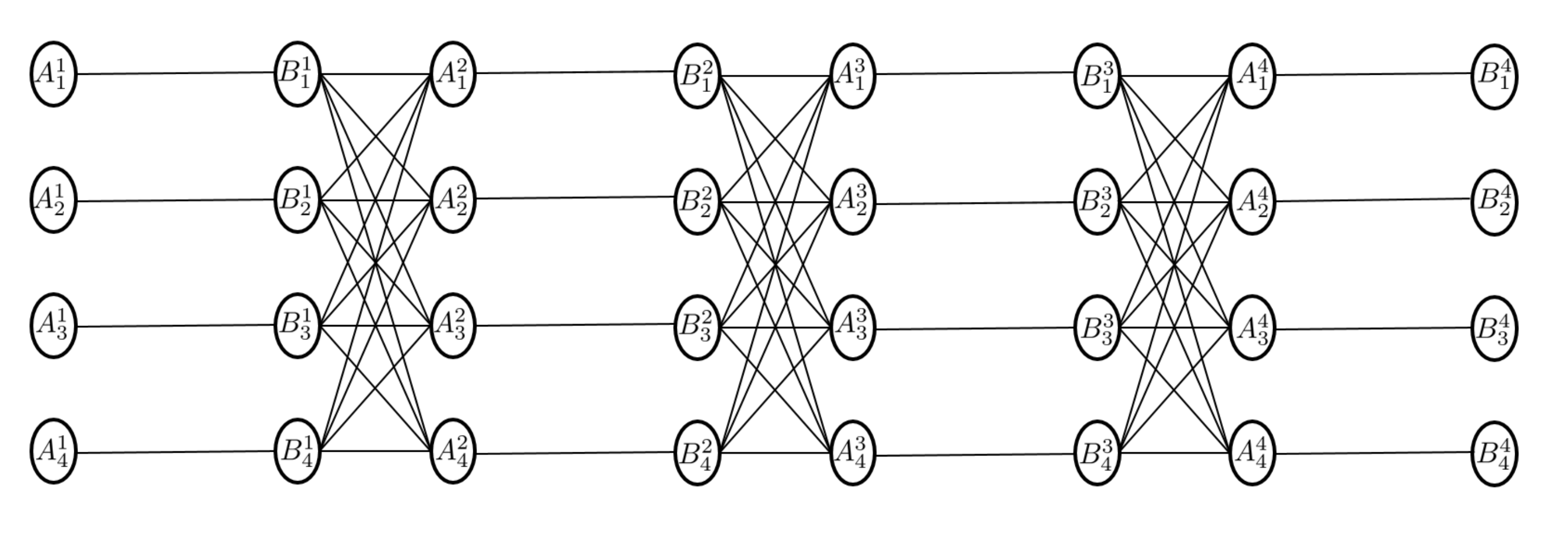}
\end{center}
\caption{$G_3$.}
\label{fig:G3}
\end{figure}

It is therefore very interesting that such a seemingly powerful method as simulated annealing can be deceived by special instances of this latter easier problem. Indeed, in a landmark 1988 paper~\cite{sasaki-1988}, Sasaki and Hajek, have considered the following family of special instances of the bipartite matching problem. Let $G_n$ denote the (undirected) graph with vertices $\bigcup_{i=0}^nA^{(i)}\cup\bigcup_{i=0}^nB^{(i)}$ where each of the $A^{(i)}$'s and $B^{(j)}$'s have cardinality $n+1$ (vertex numbering goes from $0$ to $n$), where vertex $A^{(i)}_j$ is connected to vertex $B^{(i)}_j$ and where vertex $B^{(i)}_j$ is connected to all vertices in $A^{(i+1)}$ (for $i\in\{0,\ldots,n\}$ and $j\in\{0,\ldots,n\}$). These graphs are clearly bipartite has neither two vertices in $\bigcup_{i=0}^nA^{(i)}$ nor two vertices in $\bigcup_{i=0}^nB^{(i)}$ are connected. These graphs therefore exhibit a sequential structure which alternates between sparsely and densely connected subsets of vertices, as illustrated on Figure~\ref{fig:G3} for $G_3$.

As a special case of the bipartite matching problem, the maximum cardinality matching over $G_n$ can be solved by any algorithm solving the former. Yet, it is even easier as one can easily convince oneself that a maximum matching in $G_n$ is obtained by simply selecting all the edges connecting vertices in $A^{(i)}$ to vertices in $B^{(i)}$ (for $i\in\{0,\ldots,n\})$, i.e. all the edges in the sparsely connected subsets of vertices, and that is the only way to do so. This therefore leads to a maximum matching of cardinality $(n+1)^2$.

We hence have a straightforward special case of a polynomial problem, yet the seminal result of Sasaki and Hajek states that the mathematical expectation of the number of iterations required by a large class of annealing-type algorithms to reach a maximum matching on $G_n$ is in $O(\exp(n))$. The $G_n$ family therefore provides an interesting playground to study how quantum annealing behaves on problems that are hard for simulated annealing. This is what we do, experimentally, in the sequel.

\subsection{QUBO instances}
\label{sec:qubo}


In order for our results to be fully reproducible we hereafter describe how we converted instances of the maximum matching problem into instances of the Quadratric Unconstrained Boolean Optimization (QUBO) problem which D-Wave machines require as input.

Let $G=(V,E)$ denote the (undirected) graph for which a maximum matching is desired.

We denote $x_e\in\{0,1\}$, for $e\in E$, the variable which indicates whether $e$ is in the matching.

Hence we have to maximize,
$$
\sum_{e\in E}x_e,
$$
subject to the contraints that each vertex $v$ is covered at most once, i.e. $\forall v\in V$,
\begin{equation}
\label{eq:matcons}
\sum_{e\in\Gamma(v)}x_e\leq 1,
\end{equation}
where $\Gamma(v)$, in standard graph theory notations, denotes the set of edges which have $v$ as an endpoint.

In order to turn this into a QUBO problem we have to move the above constraints into the economic function, for example in maximizing, 
\begin{eqnarray*}
&&\sum_{e\in E}x_e-\lambda\sum_{v\in V}\left(1-\sum_{e\in\Gamma(v)}x_e\right)^2\\
&=&\sum_{e\in E}x_e-\lambda\sum_{v\in V}\left(1-2\sum_{e\in\Gamma(v)}x_e+\sum_{e\in\Gamma(v)}x_e\sum_{e'\in\Gamma(v)}x_{e'}\right)\\
&=&\sum_{e\in E}x_e-\lambda|V|+\sum_{v\in V}\sum_{e\in\Gamma(v)}2\lambda x_e-\sum_{v\in V}\sum_{e\in\Gamma(v)}\sum_{e'\in\Gamma(v)}\lambda x_ex_{e'}.
\end{eqnarray*}
Dropping the constant term $-\lambda|V|$ lead to the following economic function,
$$
\sum_{e\in E}x_e+\sum_{v\in V}\sum_{e\in\Gamma(v)}2\lambda x_e-\sum_{v\in V}\sum_{e\in\Gamma(v)}\sum_{e'\in\Gamma(v)}\lambda x_ex_{e'}
$$

Yet we have to reorganize a little to build a proper QUBO matrix. Let $e=(v,w)$, variable $x_e$ has coefficient 1 in the first term, $2\lambda$ in the second term (for $v$) then $2\lambda$ again in the second term (for $w$) then $-\lambda$ in the third term (for $v$ and $e'=e$) and another $-\lambda$ again in the third term (for $w$ and $e'=e$). Hence, the diagonal terms of the QUBO matrix are, 
$$
Q_{ee}=1+4\lambda-2\lambda=1+2\lambda.
$$
Then, if two distinct edges $e$ and $e'$ share a common vertex, the product of variables $x_ex_{e'}$ has coefficient $-\lambda$, in the third term, when $v$ corresponds to the vertex shared by the two edges, and this is so twice. So, for $e\neq e'$,
$$
Q_{ee'}=\left\{\begin{array}{rl}
  -2\lambda & \text{if $e\cap e'\neq\emptyset$},\\
  0 & \text{otherwise}.
\end{array}\right.
$$
Taking $\lambda=\card{E}$\footnote{As $\card{E}$ is clearly an upper bound for the cost of any matching, any solution which violates at least one of the constraints \eqref{eq:matcons} cannot be optimal.}, for example for $G_1$, we thus obtain the 8 variables QUBO defined by the following matrix,
$$
\left(
\begin{array}{c|rrrrrrrr}
  &   0 &   1 &   2 &   3 &   4 &   5 &   6 &   7 \\
\hline
0 &  17 &   0 & -16 & -16 &   0 &   0 &   0 &   0 \\
1 &   0 &  17 &   0 &   0 & -16 & -16 &   0 &   0 \\
2 &   0 &   0 &  17 & -16 & -16 &   0 & -16 &   0 \\
3 &   0 &   0 &   0 &  17 &   0 & -16 &   0 & -16 \\
4 &   0 &   0 &   0 &   0 &  17 & -16 & -16 &   0 \\
5 &   0 &   0 &   0 &   0 &   0 &  17 &   0 & -16 \\
6 &   0 &   0 &   0 &   0 &   0 &   0 &  17 &   0 \\
7 &   0 &   0 &   0 &   0 &   0 &   0 &   0 &  17 
\end{array}
\right),
$$
for which a maximum matching has cost 68 and the second best solutions has cost 53 and the worst one (which consist in selecting all edges) has cost -56.

\section{Experimental results}
\label{sec:exp}

\subsection{Concrete implementation on a D-Wave}
In this section, we detail the steps that we have followed to concretely map and solve the QUBO instances associated to $G_n$, $n \in \{1,2,3,4\}$, on a DW2X operated by the University of South California.

Unfortunately (yet unsurprisingly), the QUBO matrices defined in the previous section are not directly mappable on the Chimera interconnection topology and, thus, we need to resort to qubit duplication i.e., use \emph{several} physical qubits to represent \emph{one} problem variable (or ``logical qubit''). Fortunately, the D-Wave software pipeline automates this duplication process. Yet, this need for duplication (or equivalently the sparsity of the Chimera interconnection topology) severely limits the size of the instances we were able to map on the device and we had to stop at $G_4$ which 125 variables required using 951 of the 1098 available qubits. Table \ref{tab:res1} provides the number of qubits required for each of our four instances.


\begin{table}[htbp]
\begin{center}
\begin{tabular}{|c|cccc|}
\hline
 & \#var. & \#qubits & average dup. & max. dup. \\
\hline
$G_1$ &8  &16  &2.0  & 6\\
$G_2$ &27  &100  &3.7   & 6 \\
$G_3$ & 64  &431  &6.7  & 18 \\
$G_4$ & 125  &951  &7.6 &  18 \\
\hline 
\end{tabular}
\end{center}
\caption{Number of qubits required to handle the QUBO instances associated to $G_1$, $G_2$, $G_3$ and $G_4$. See text.}
\label{tab:res1}
\end{table}

\begin{figure}[htbp]
\begin{center}
\includegraphics[width=\textwidth]{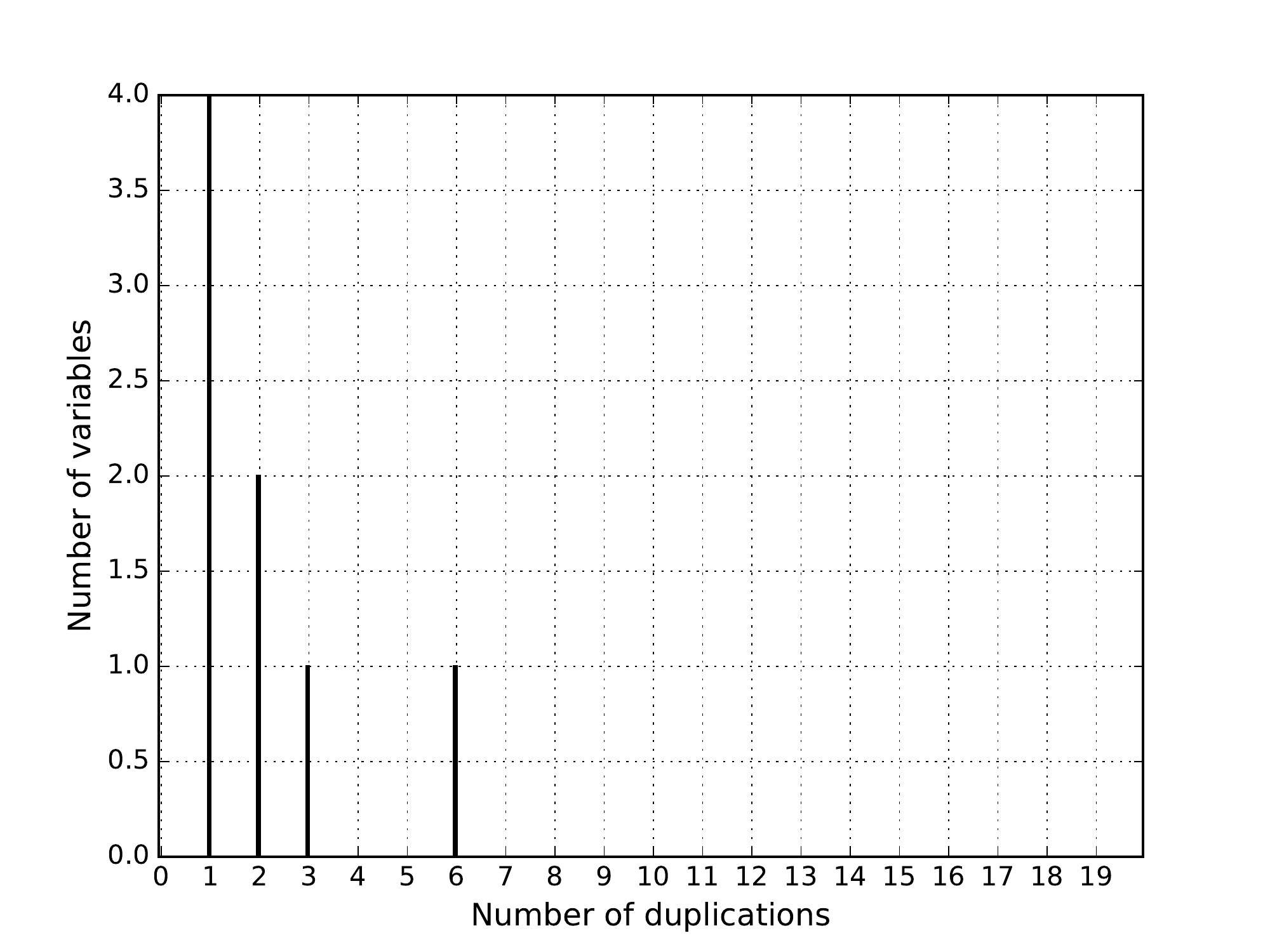}
\end{center}
\caption{Histogram for the number of duplications for $G_1$. The maximum duplication is 6 qubits.}
\label{fig:duplicationg1}
\end{figure}

\begin{figure}[htbp]
\begin{center}
\includegraphics[width=\textwidth]{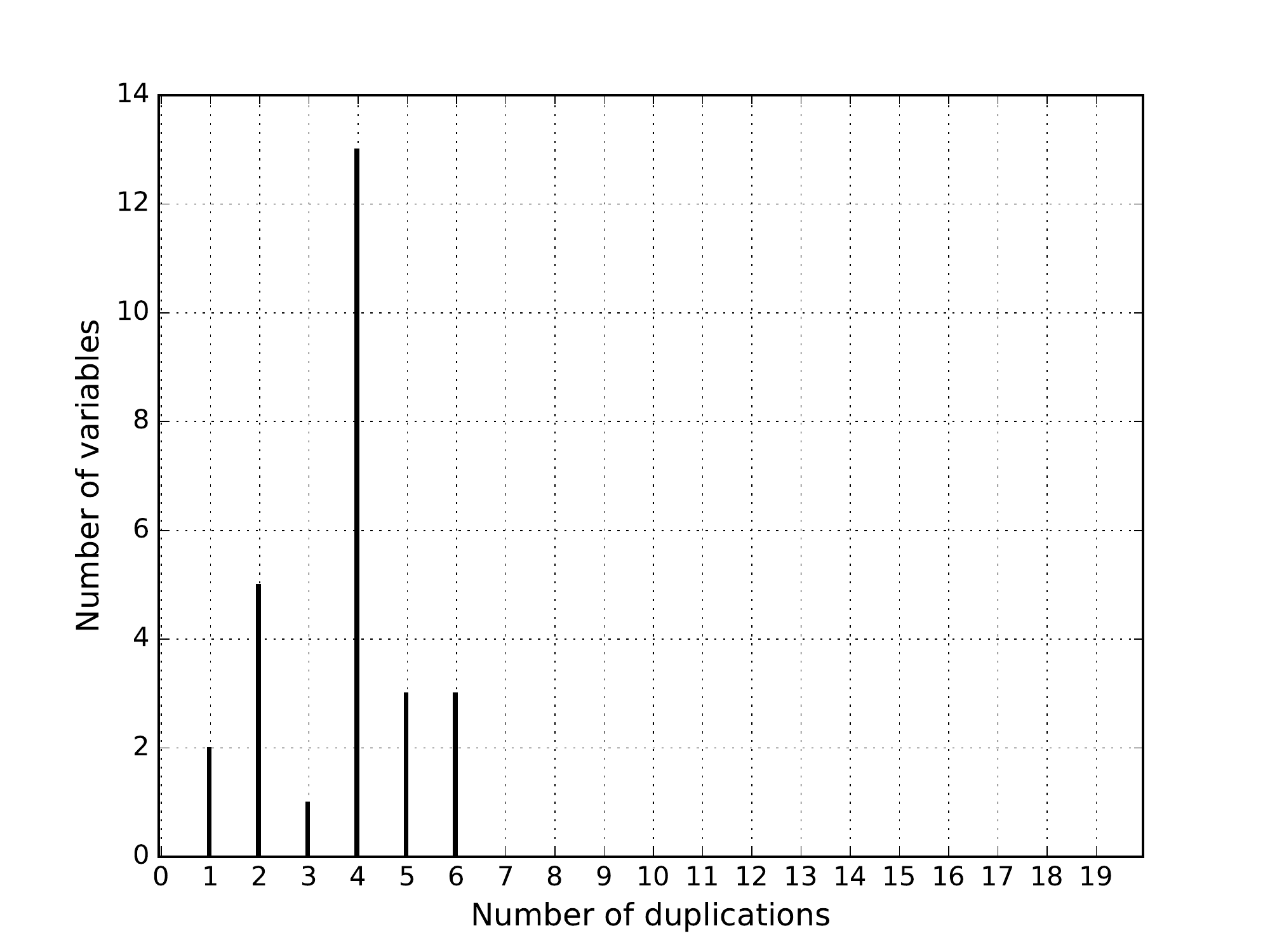}
\end{center}
\caption{Histogram for the number of duplications for $G_2$. The maximum duplication is 6 qubits.}
\label{fig:duplicationg2}
\end{figure}

\begin{figure}[htbp]
\begin{center}
\includegraphics[width=\textwidth]{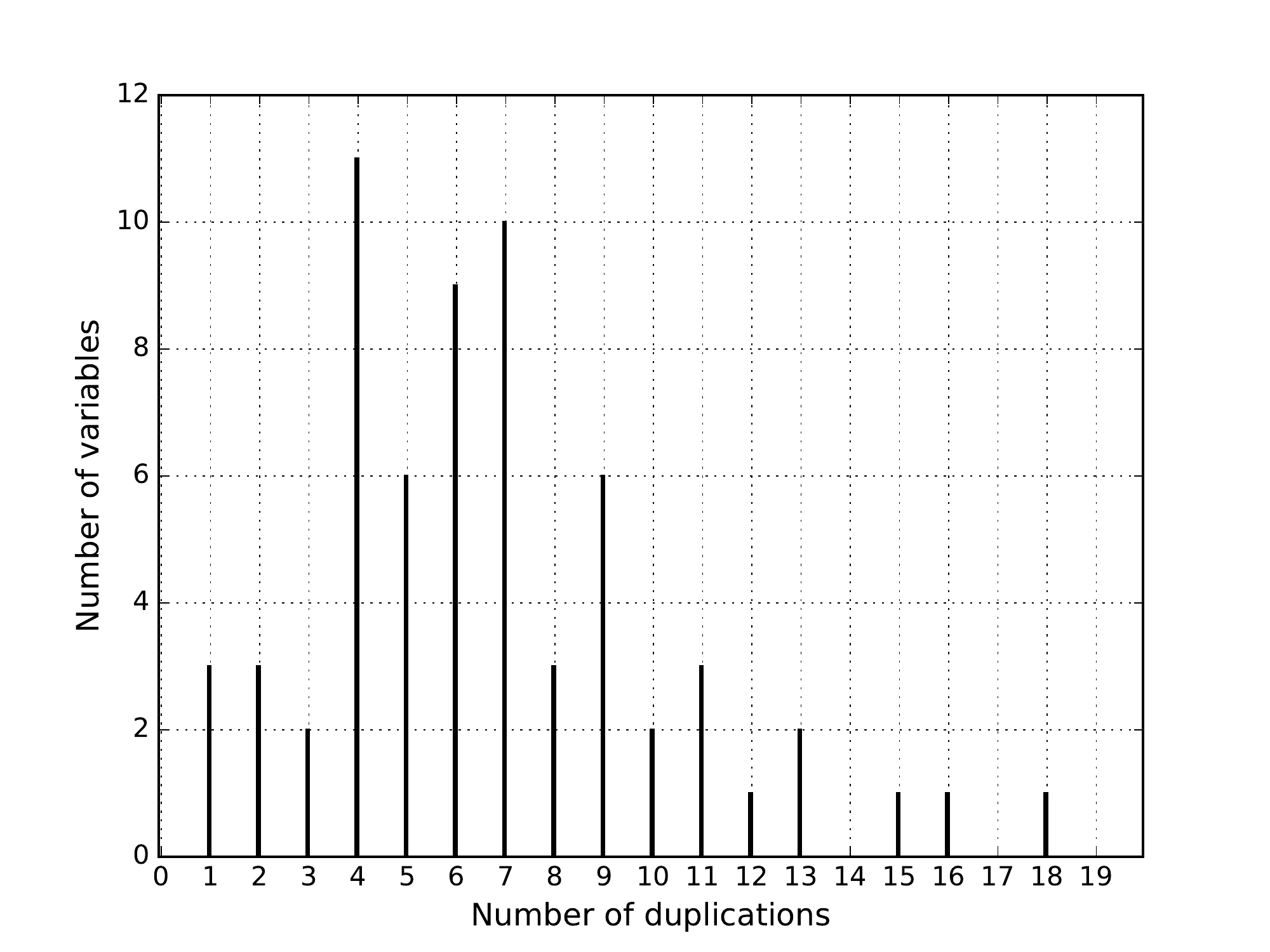}
\end{center}
\caption{Histogram for the number of duplications for $G_3$. The maximum duplication is 18 qubits.}
\label{fig:duplicationg3}
\end{figure}

\begin{figure}[htbp]
\begin{center}
\includegraphics[width=\textwidth]{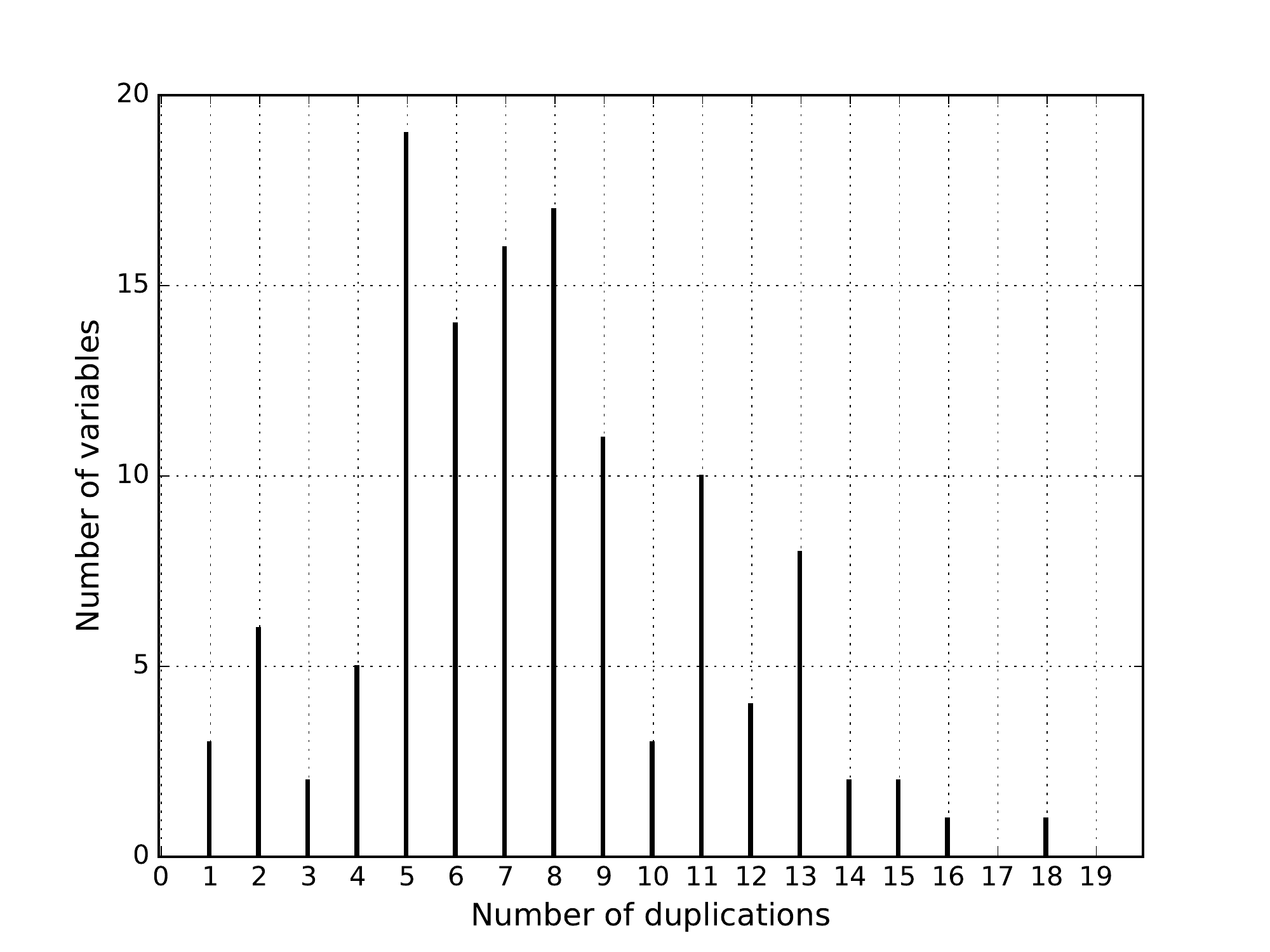}
\end{center}
\caption{Histogram for the number of duplications for $G_4$. The maximum duplication is 18 qubits.}
\label{fig:duplicationg4}
\end{figure}

Additionally, Figures \ref{fig:duplicationg1}, \ref{fig:duplicationg2}, \ref{fig:duplicationg3} and \ref{fig:duplicationg4} provides the histogram of the number of duplications for $G_1$, $G_2$, $G_3$ and $G_4$.

Eventually, qubit duplication leads to an expanded QUBO with more variables and an economic function which includes an additional set of penalty constraints to favor solutions in which qubits representing the same variable indeed end up with the same value. More precisely, each pair of distinct qubits $q$ and $q'$ (associated to the same QUBO variable) adds a penalty term of the form
$$
\varphi q(1-q')
$$
where the penalty constant $\varphi$ is (user) chosen as minus the cost of the worst possible solution to the initial QUBO which is obtained for a vector filled with ones (i.e., a solution that selects all edges of the graph and which therefore maximizes the highly-penalized violations of the cardinality contraints). This therefore guarantees that a solution which violates at least one of these consistency constraints cannont be optimal (please note that we have switched from a maximization problem in Sect. \ref{sec:qubo} to a minimization problem as required by the machine)

Lastly, as qubit duplication leads to an expanded QUBO which support graph is trivially isomorphic to the Chimera topology, it can be mapped on the device after a renormalization of its coefficients to ensure that the diagonal terms of $Q$ are in $[-2,2]$ and the others in $[-1,1]$.

\subsection{Results summary}

This section reports on the experiments we have been able to perform on instances of the previous QUBO problems. As already emphasized, due to the sparsity of the qubit interconnection topology, our QUBO instances were not directly mappable on the D-Wave machine and we had to resort to qubit duplications (whereby one problem variable is represented by several qubits on the D-Wave, bound together to end up with the same value at the end of the annealing process). This need for qubit duplication limited us to $G_4$ which, with 125 binary variables, already leads to a combinatorial problem of non trivial size. Yet, to solve it, we had to mobilize about $87\%$ of the 1098 qubits of the machine. The results below have been obtained by running 10000 times the quantum annealer with a 20 $\mu$s annealing time (although we also experimented with 200 and 2000 $\mu$s, which did not appear to affect the results significantly). Table \ref{tab:res} summarizes key statistics of the obtained results. The following paragraphs discuss each instance in greater details.

\begin{table}[htbp]
\begin{center}
\begin{tabular}{|c|cccccc|}
\hline
 & opt. & best sol. & worst sol. & mean & median & stdev \\
\hline
$G_1$  &-68  &-68  &-6  &-67.4  &-68 &3.2  \\
$G_2$ &-495  &-495  &-89  &-402.9  &-388 &47.8  \\
$G_3$ &-2064  &-1809  &-549  &-1460.8  &-1549 &136.4  \\
$G_4$ &-6275 &-5524  &-2109  &-4492.4  &-4525 &391.8  \\
\hline 
\end{tabular}
\end{center}
\caption{Experimental results summary on $G_1$, $G_2$, $G_3$, $G_4$. See text.}
\label{tab:res}
\end{table}

\subsection{Instances solutions}

\begin{figure}[htbp]
\begin{center}
\includegraphics[height=1.3\textwidth]{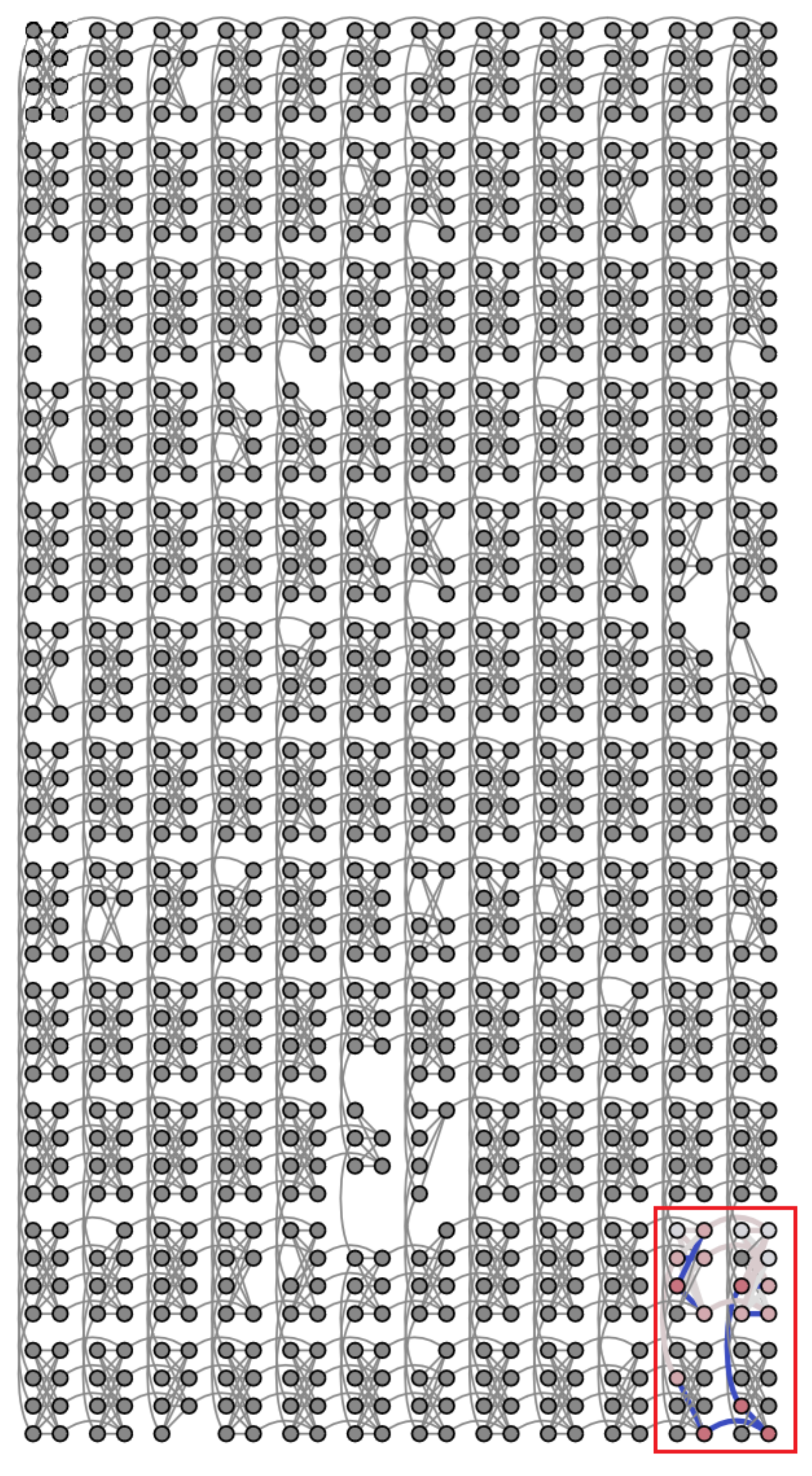}
\end{center}
\caption{Mapping of the QUBO instance associated to $G_1$ on our DW2X with variable 0 being mapped to qubits $\{1040\}$, 1 to $\{1048\}$, 2 to $\{1053\}$, 3 to $\{1055, 1051\}$, 4 to $\{1041, 1045\}$, 5 to $\{1044, 1042, 1047\}$, 6 to  $\{1137, 1143, 1054, 1151, 1050, 1146\}$ and 7 to $\{1052\}$.}
\label{fig:G1:map}
\end{figure}
\begin{figure}[htbp]
\begin{center}
\includegraphics[height=0.34\textwidth]{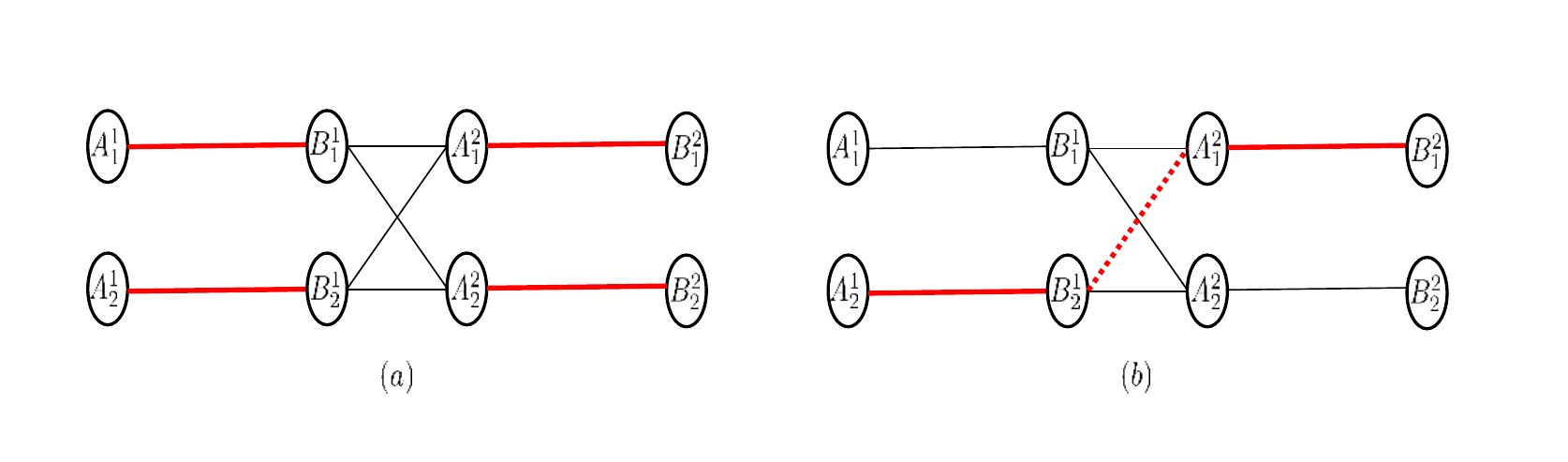}
\end{center}
\caption{Graphical representation of the solutions in Table~\ref{tab:dupg1}. Dotted lines indicates duplication inconsistencies.}
\label{fig:G1bw}
\end{figure}
\begin{figure}[htbp]
\begin{center}
\includegraphics[width=\textwidth]{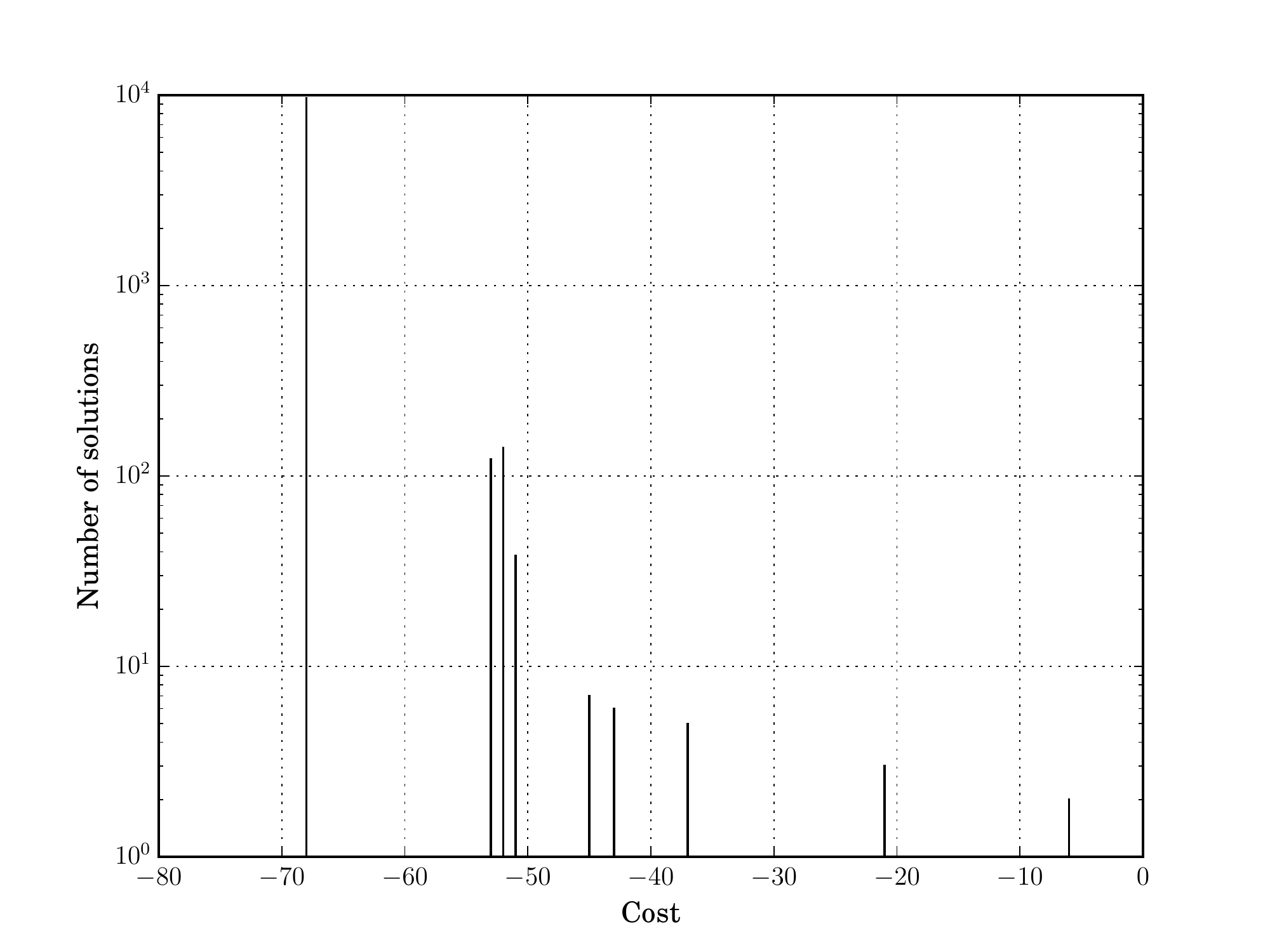}
\end{center}
\caption{Histogram of the economic function over 10000 annealing runs on $G_1$.}
\label{fig:G1:histo}
\end{figure}
\begin{figure}[htbp]
\begin{center}
\includegraphics[width=\textwidth]{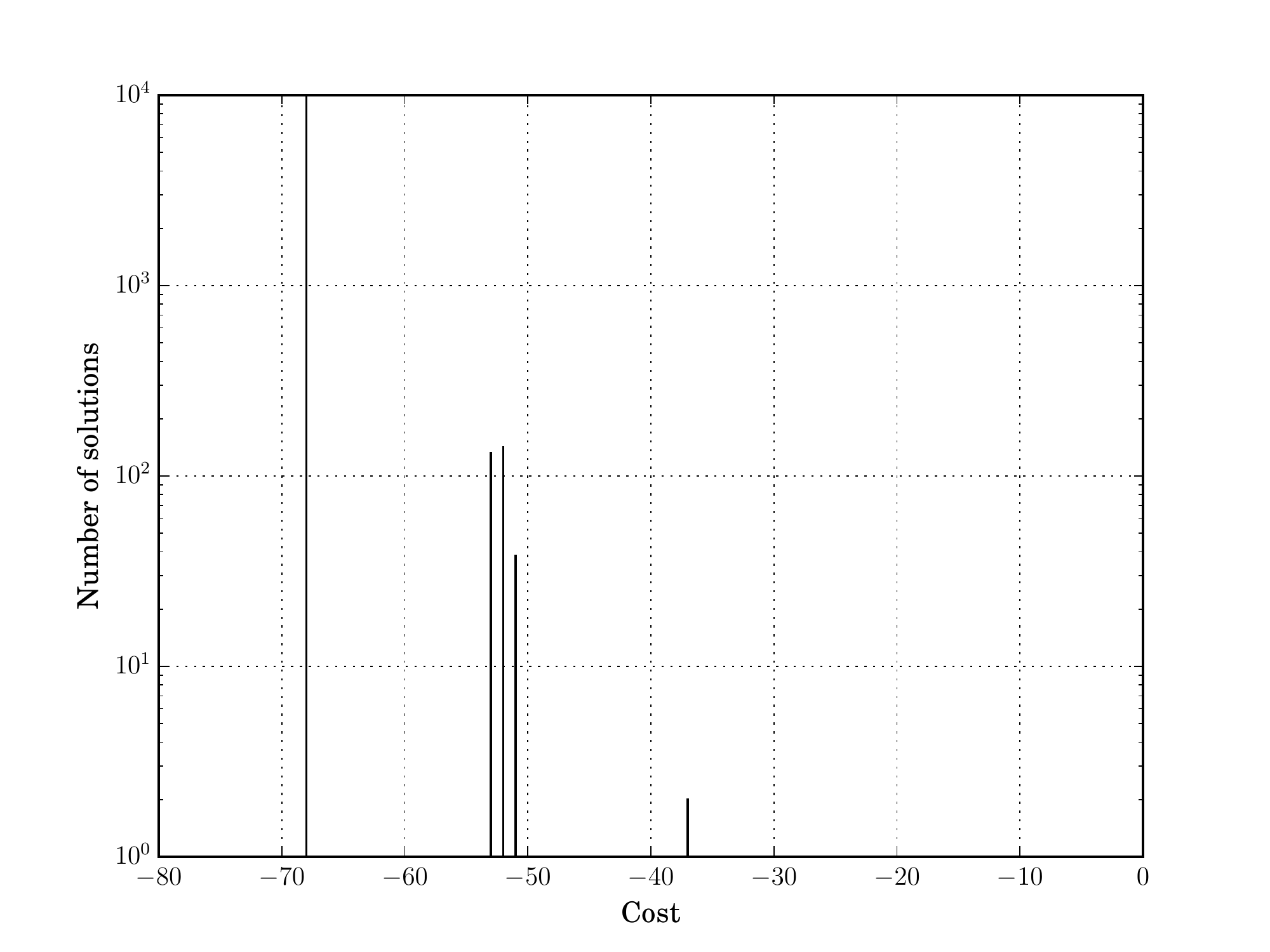}
\end{center}
\caption{Histogram of the economic function over 10000 annealing runs on $G_1$ (with duplication inconsistencies fixed by majority voting).}
\label{fig:G1:correct}
\end{figure}

\paragraph{\textbf{$G_1$.}} This instance leads to a graph with 8 vertices, 8 edges and then (before duplication) to a QUBO with 8 variables and 12 nonzero nondiagonal coefficients\footnote{In the Chimera topology the diagonal coefficient are not constraining as there is no limitation on the qubits autocouplings.}. Mapping this QUBO on the D-Wave machine required 16 qubits as shown on Figure~\ref{fig:G1:map}.  Over 10000 runs, the optimal solution was obtained 9673 times. Table \ref{tab:dupg1} and Figure \ref{fig:G1bw} illustrate the best (with a cost of $-68$) and worst solutions (with a cost of $-6$) obtained for $G_1$ (the median solution is identical to the best one for $G_1$). Interestingly, the worst solution obtained violates duplication consistency as all the 6 qubits representing variable 6 do not have the same value (6 of them are 0, so in that particular case, rounding the solution by means of majority voting gives the optimal solution). Figure~\ref{fig:G1:histo} shows the histogram of the economic function as outputted by the D-Wave (but renormalized) for the 10000 annealing runs we performed. Additionally, since some of the solutions obtained by the D-Wave are inconsistent with respect to duplication, Figure \ref{fig:G1:correct} shows the histogram of the economic function for the solutions in which duplication inconsistencies were fixed by majority voting.

\begin{table}[htbp]
\begin{center}
\begin{tabular}{|c|c||c||c|}
\hline
qubits & variable & best & worst\\
\hline
1040 &0 &1 &1\\
1041 &4 &0 &0\\
1042 &5 &0 &0\\
1044 &5 &0 &0 \\
1045 &4 &0 &0\\
1047 &5 &0 &0\\
1048 &1 &1 &1\\
1050 &6 &0 &0\\
1051 &3 &1 &1\\
1052 &7 &0 &0\\
1053 &2 &1 &1\\
1054 &6 &0 &0\\
1055 &3 &1 &1\\
1137 &6 &0 &0\\
1143 &6 &0 &1\\
1146 &6 &0 &0\\
1151 &6 &0 &1\\
\hline 
\end{tabular}
\end{center}
\caption{Selection of solutions (best and worst) for the expanded QUBO associated to $G_1$. The best and median solution are identical (and optimal) in the case of $G_1$ and have cost $-68$. The worst solution has cost $-6$. In that latter solution, the 6 qubits representing variable $6$ do not have the same values.}
\label{tab:dupg1}
\end{table}

\begin{figure}[htbp]
\begin{center}
\includegraphics[height=1.3\textwidth]{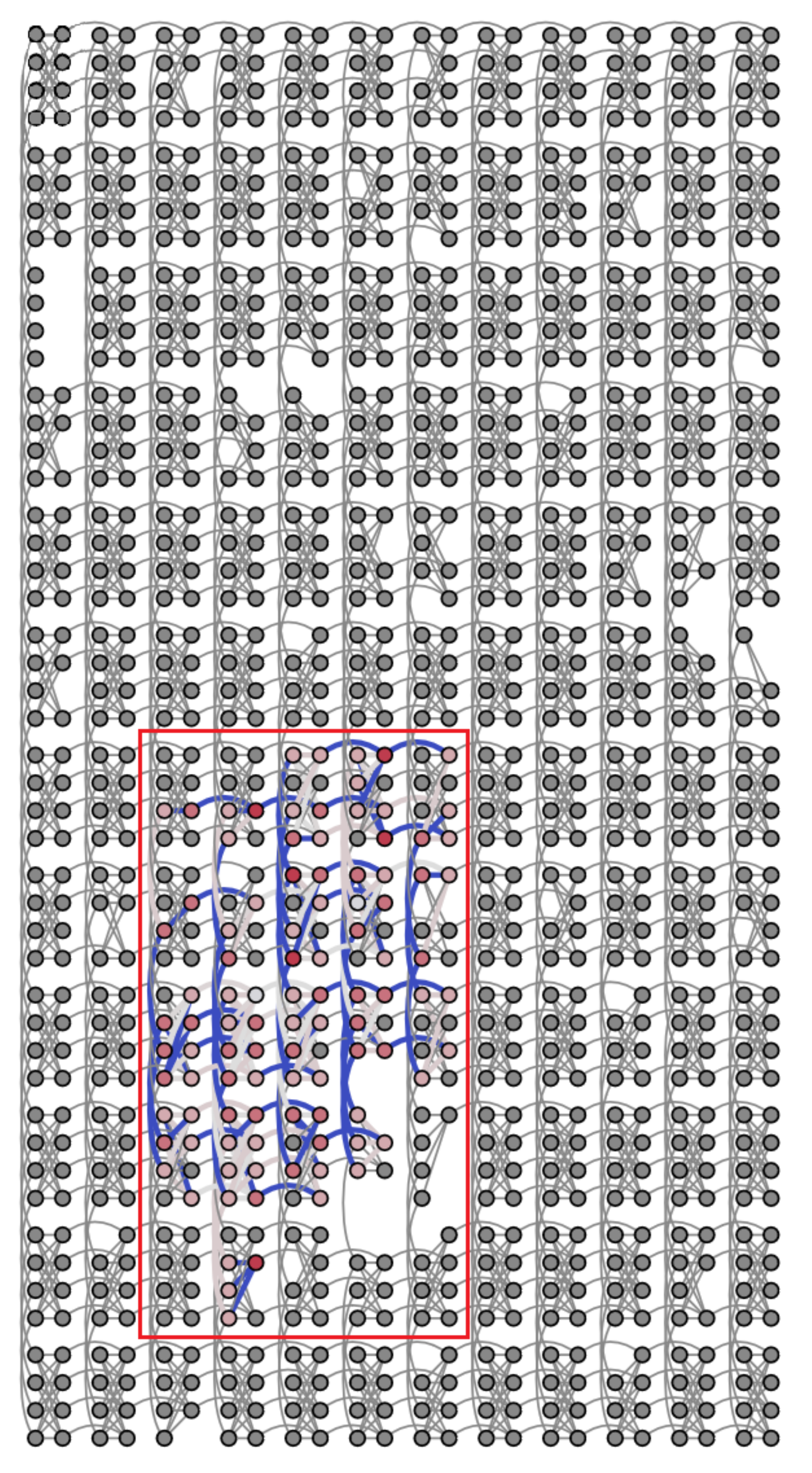}
\end{center}
\caption{Mapping of the QUBO instance associated to $G_2$ on the D-Wave 2X.}
\label{fig:G2:map}
\end{figure}

\begin{figure}[htbp]
\begin{center}
\includegraphics[height=\textwidth]{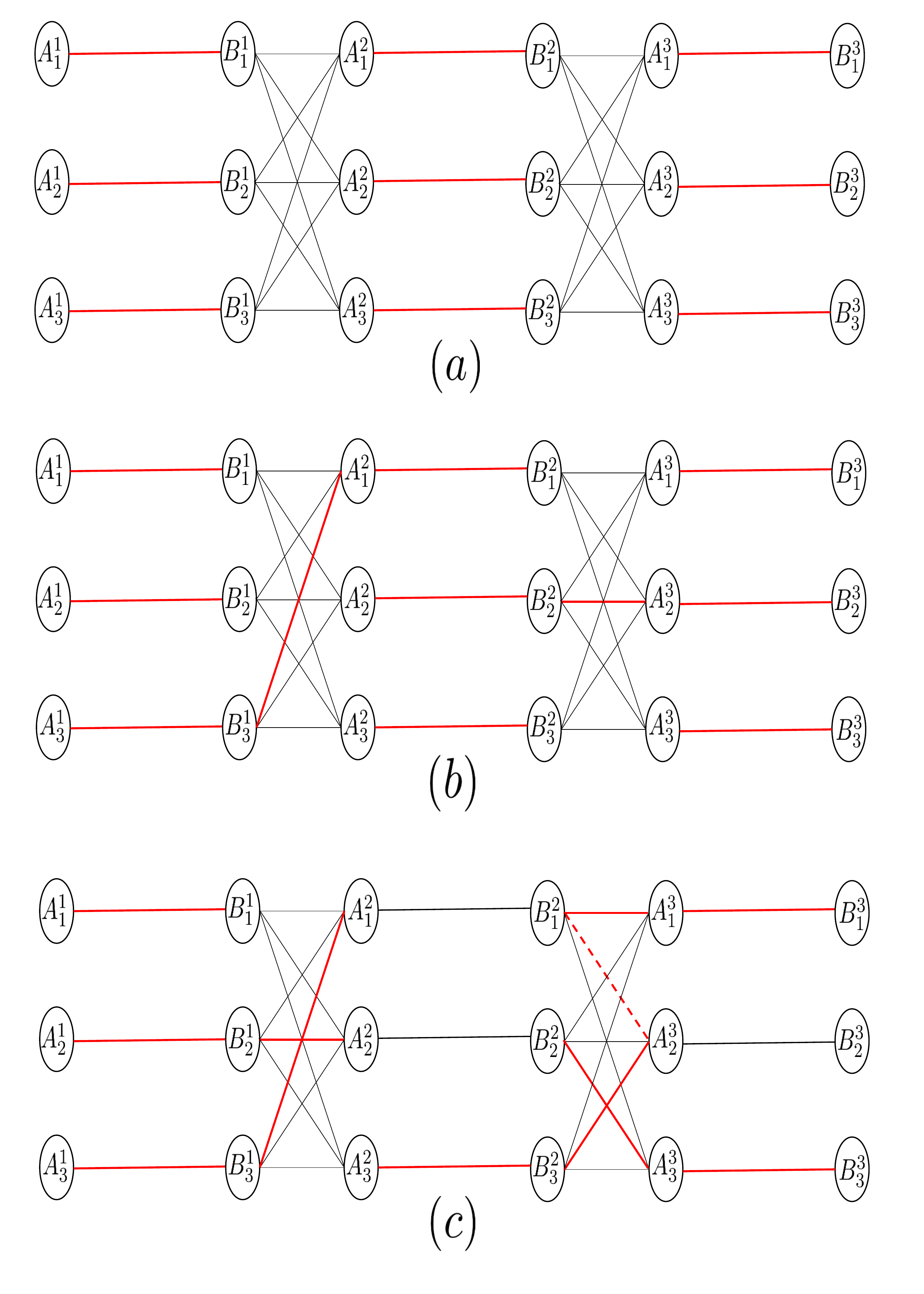}
\end{center}
\caption{Graphic representation of the best (a), median (b) and worst (c) solution obtained for $G_2$. Dotted lines represent qubit duplication inconsistencies. See text.}
\label{fig:G2bw}
\end{figure}

\begin{figure}[htbp]
\begin{center}
\includegraphics[width=\textwidth]{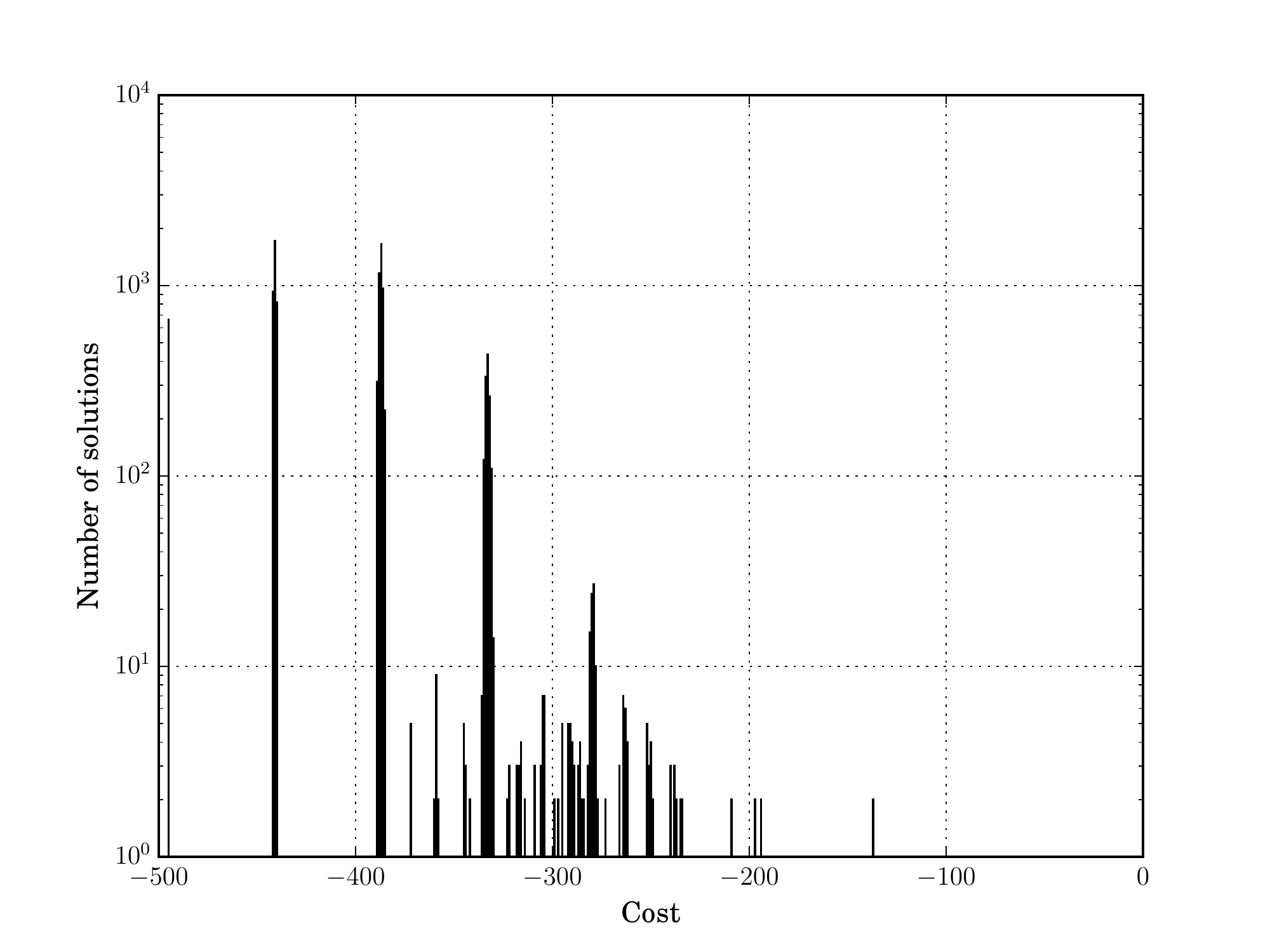}
\end{center}
\caption{Histogram of the economic function over 10000 annealing runs on $G_2$.}
\label{fig:G2:histo}
\end{figure}

\begin{figure}[htbp]
\begin{center}
\includegraphics[width=\textwidth]{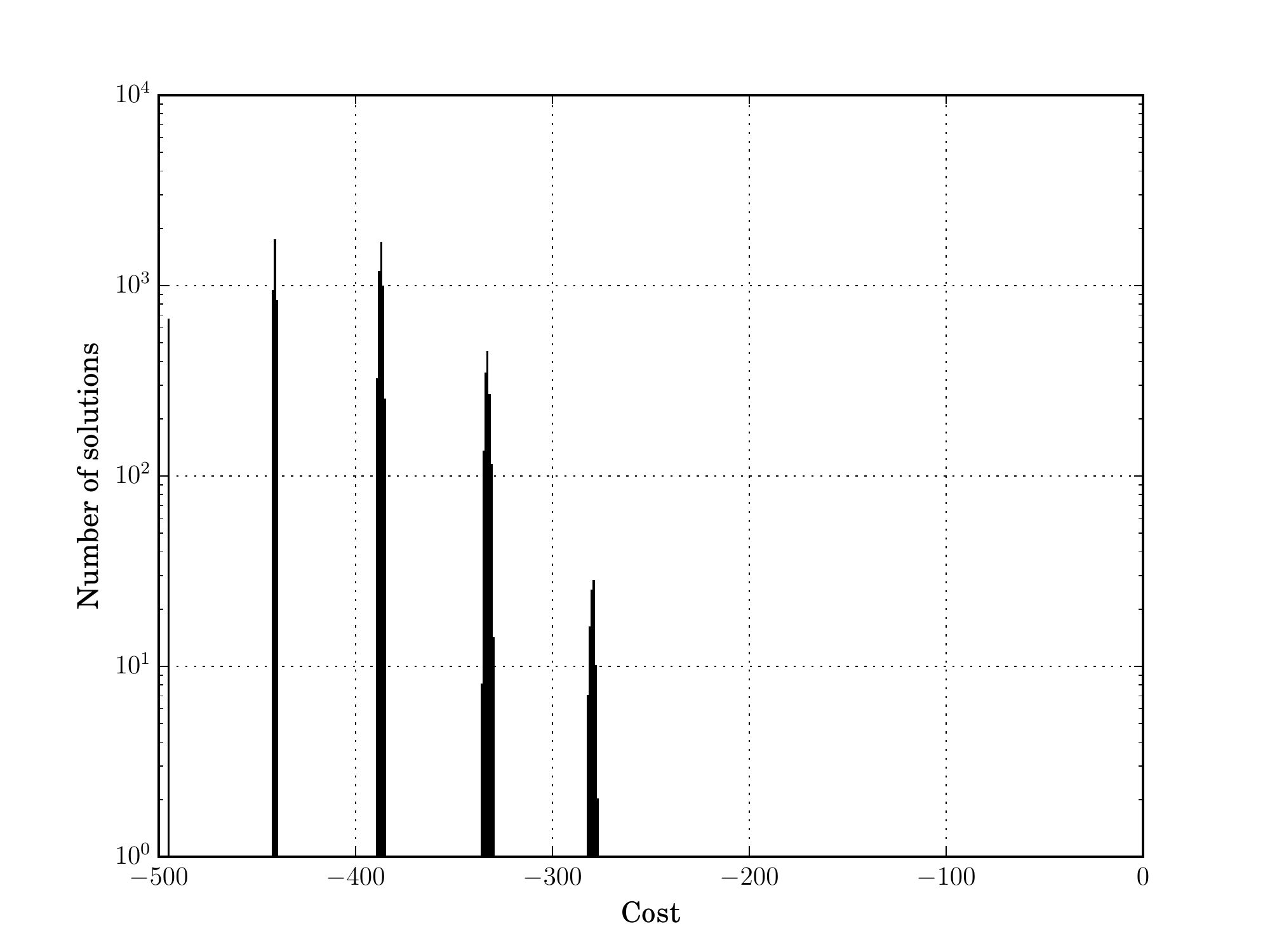}
\end{center}
\caption{Histogram of the economic function over 10000 annealing runs on $G_2$ (with duplication inconsistencies fixed by majority voting).}
\label{fig:G2:correct}
\end{figure}

\paragraph{\textbf{$G_2$.}} This instance leads to a graph with 18 vertices, 27 edges and then to a QUBO with 27 variables and 72 nonzero nondiagonal coefficients. Mapping this QUBO on the D-Wave machine required 100 qubits as shown on Figure~\ref{fig:G2:map}.  Over 10000 runs the optimal solution was obtained only 662 times (i.e., a ~6\% hitting probability). Figure \ref{fig:G2bw} provides graphic representations of the best, median and worst solutions obtained (respectively with cost $-495$, $-389$ and $-89$). Although the best solution obtained is optimal, the median solution does not lead to a valid matching since four vertices are covered 3 times\footnote{Fixing this would require a postprocessing step to produce valid matchings. Of course this is of no relevance for a polynomial problem, but such a postprocessing would thus be required when operationally using a D-Wave for solving non artificial problems.}. As for $G_1$, we also observe that the worst solutions has duplication consistency issues. Figure~\ref{fig:G2:histo} shows the histogram of the economic function as outputted by the D-Wave (but renormalized) for the 10000 annealing runs we performed. Additionally, since some of these solutions are inconsistent with respect to duplication, Figure~\ref{fig:G2:correct} shows the histogram of the economic function for the solutions in which duplication inconsistencies were fixed by majority voting (resulting in a marginal left shift of the average solution cost from -402.9 to -404.8, the median being unchanged).

\begin{figure}[htbp]
\begin{center}
\includegraphics[height=1.3\textwidth]{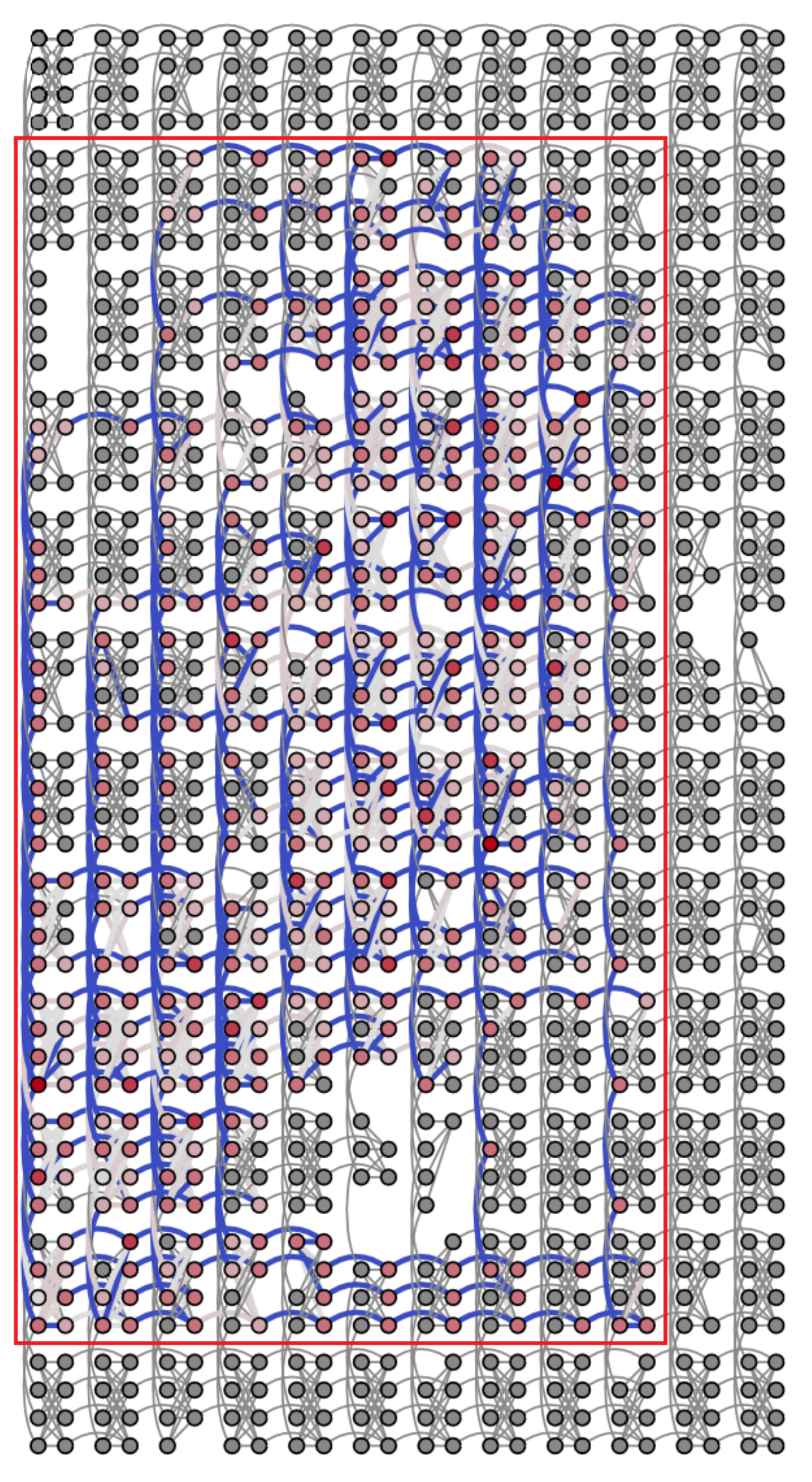}
\end{center}
\caption{Mapping of the QUBO instance associated to $G_3$ on our D-Wave 2X.}
\label{fig:G3:map}
\end{figure}

\begin{figure}[htbp]
\begin{center}
\includegraphics[height=1\textwidth]{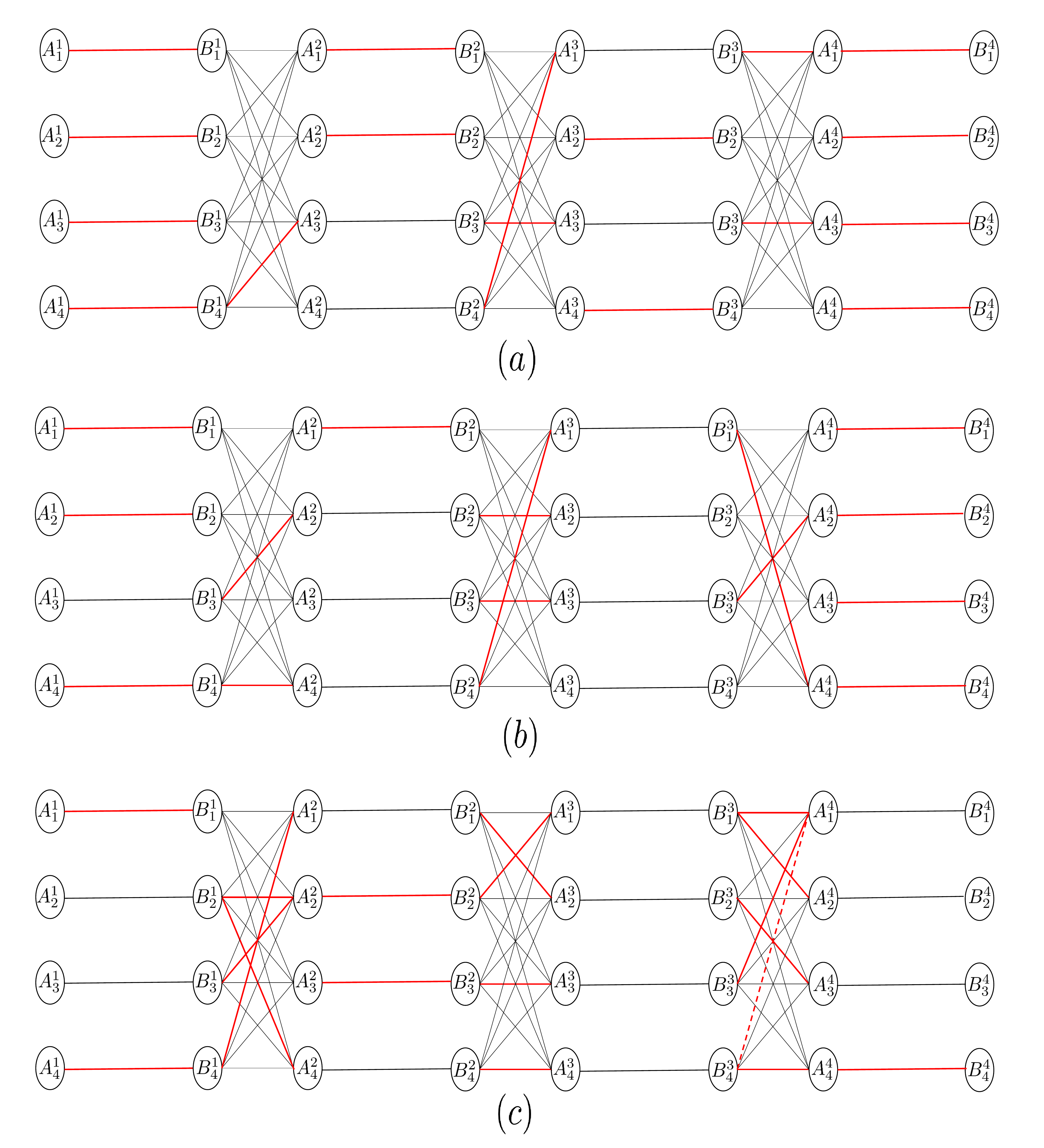}
\end{center}
\caption{Graphic representation of the best (a), median (b) and worst (c) solution obtained for $G_3$. Dotted lines represent qubit duplication inconsistencies. See text.}
\label{fig:G3bw}
\end{figure}

\begin{figure}[htbp]
\begin{center}
\includegraphics[width=\textwidth]{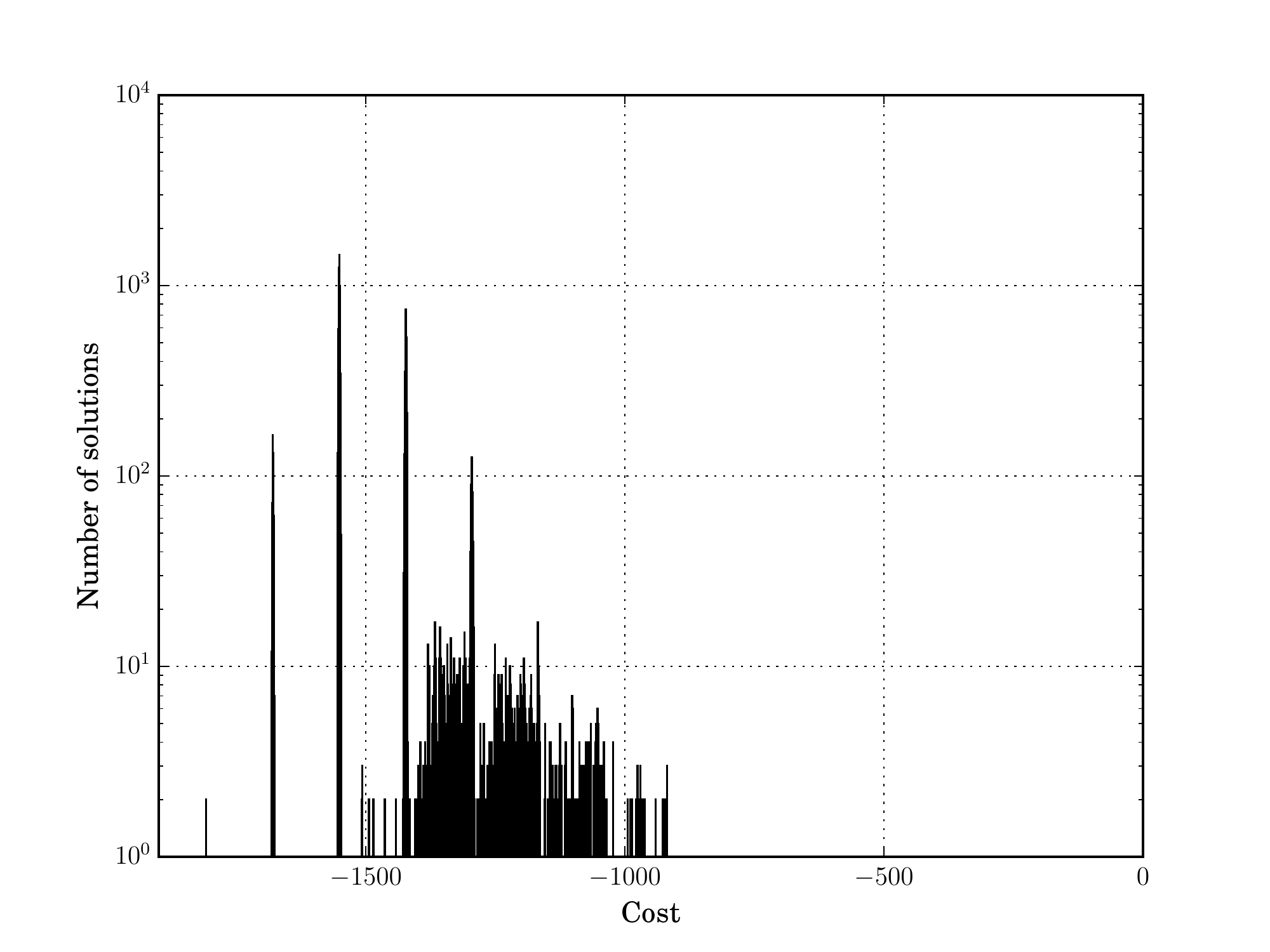}
\end{center}
\caption{Histogram of the economic function over 10000 annealing runs on $G_3$.}
\label{fig:G3:histo}
\end{figure}

\begin{figure}[htbp]
\begin{center}
\includegraphics[width=\textwidth]{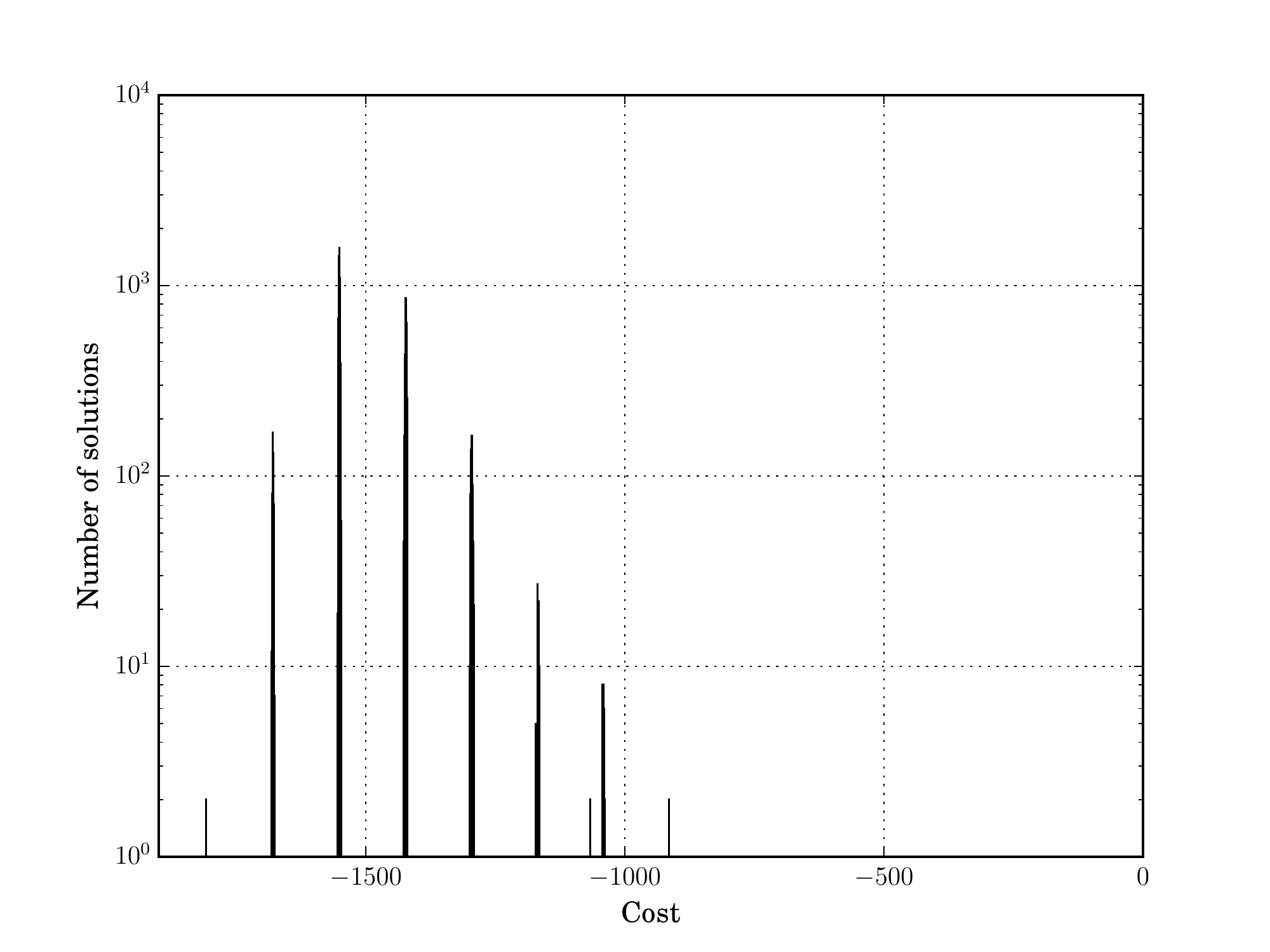}
\end{center}
\caption{Histogram of the economic function over 10000 annealing runs on $G_3$ (with duplication inconsistencies fixed by majority voting)}
\label{fig:G3:correct}
\end{figure}

\paragraph{\textbf{$G_3$.}} This instance leads to a graph with 32 vertices, 64 edges and then to a QUBO with 64 variables and 240 nonzero nondiagonal coefficients. Mapping this QUBO on the D-Wave machine required 431 qubits ($39\%$ of the available qubits) as shown on Figure~\ref{fig:G3:map}. Over 10000 runs the optimal solution was never obtained. Still, Figure \ref{fig:G3bw} provides graphic representations of the best, median and worst solutions obtained (respectively with cost $-1809$, $-1551$ and $-549$). For $G_3$, the optimum value is $-2064$, thus the best solution obtained is around 15\% far off (median cost is 25\%). Furthermore, neither the best nor the median solution lead to valid matchings since in both, some vertices are covered several times. We also observe that the worst solution has duplication consistency issues. Figure~\ref{fig:G3:histo} shows the histogram of the economic function as outputted by the D-Wave (but renormalized) for the 10000 annealing runs we performed. Additionally, since some of these solutions are inconsistent with respect to duplication, Figure \ref{fig:G3:correct} shows the histogram of the economic function for the solutions in which duplication inconsistencies were fixed by majority voting (thus left shifting the average cost from -1460.8 to -1491.8 and the median cost from -1549 to -1550 which is marginal).

\begin{figure}[htbp]
\begin{center}
\includegraphics[height=1.4\textwidth]{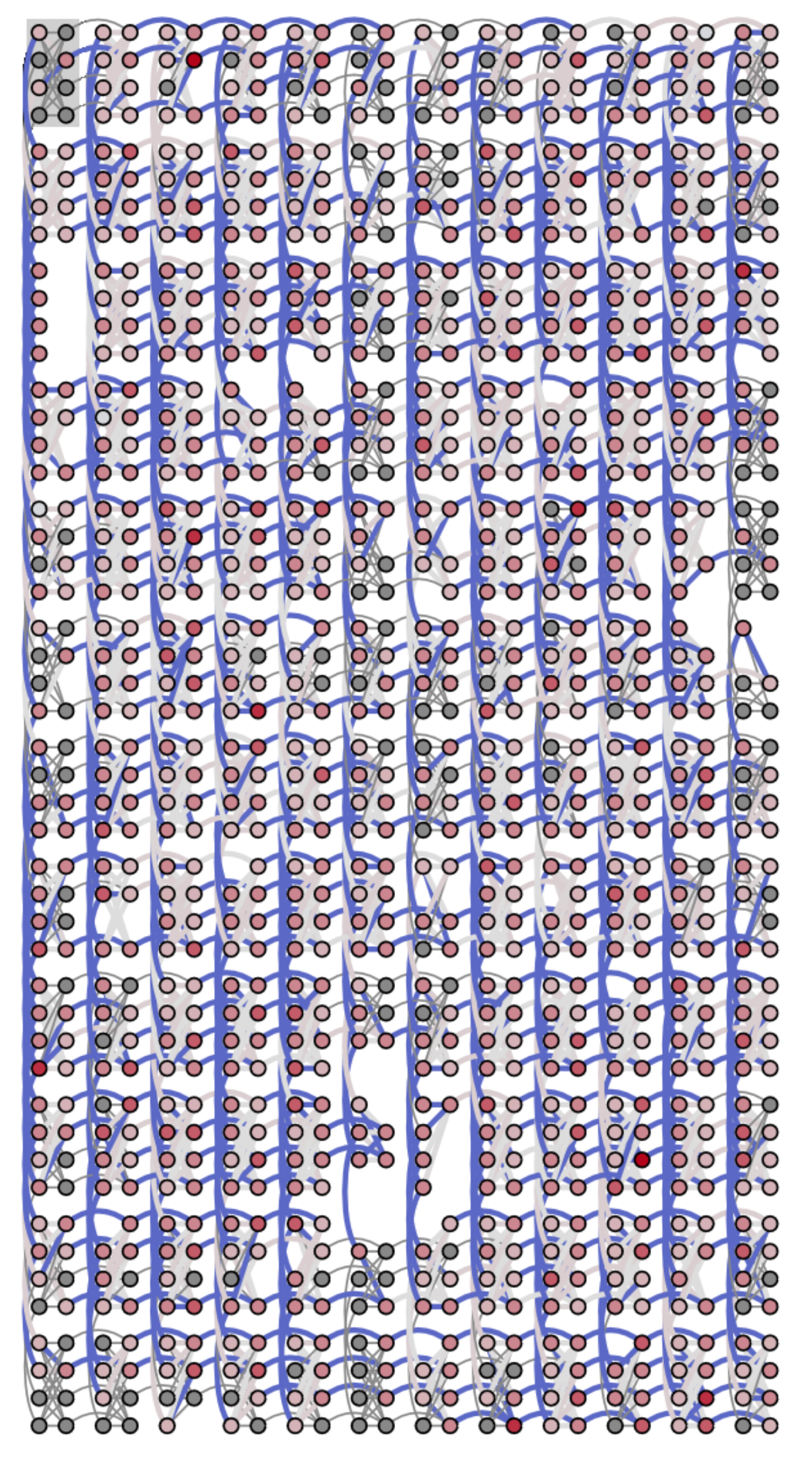}
\end{center}
\caption{Mapping of the QUBO instance associated to $G_4$ on our D-Wave 2X.}
\label{fig:G4:map}
\end{figure}

\begin{figure}[htbp]
\begin{center}
\includegraphics[width=\textwidth]{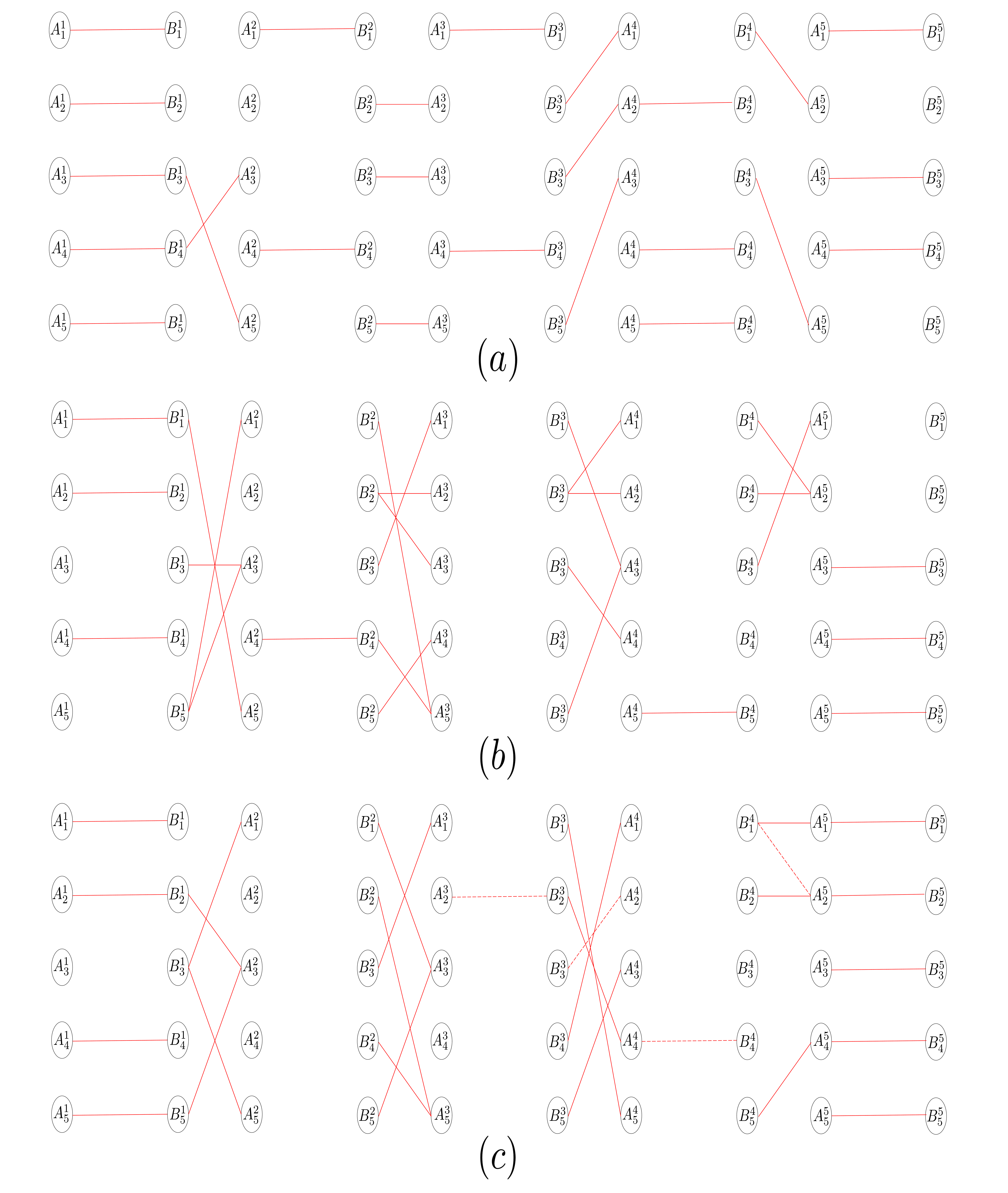}
\end{center}
\caption{Graphic representation of the best (a), median (b) and worst (c) solution obtained for $G_4$. Dotted lines represent qubit duplication inconsistencies. See text.}
\label{fig:G4bw}
\end{figure}

\begin{figure}[htbp]
\begin{center}
\includegraphics[width=\textwidth]{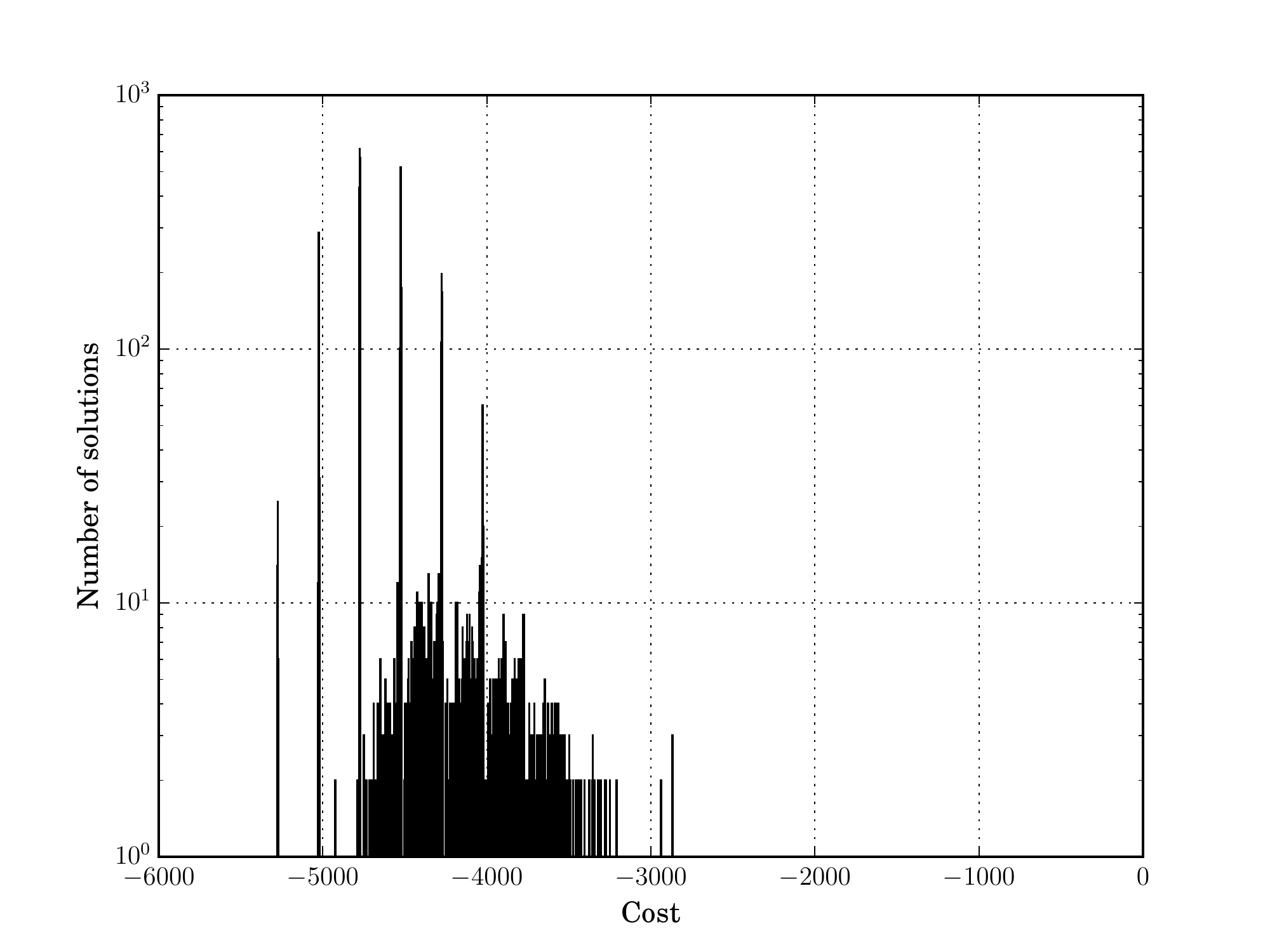}
\end{center}
\caption{Histogram of the economic function over 10000 annealing runs on $G_4$.}
\label{fig:G4:histo}
\end{figure}

\begin{figure}[htbp]
\begin{center}
\includegraphics[width=\textwidth]{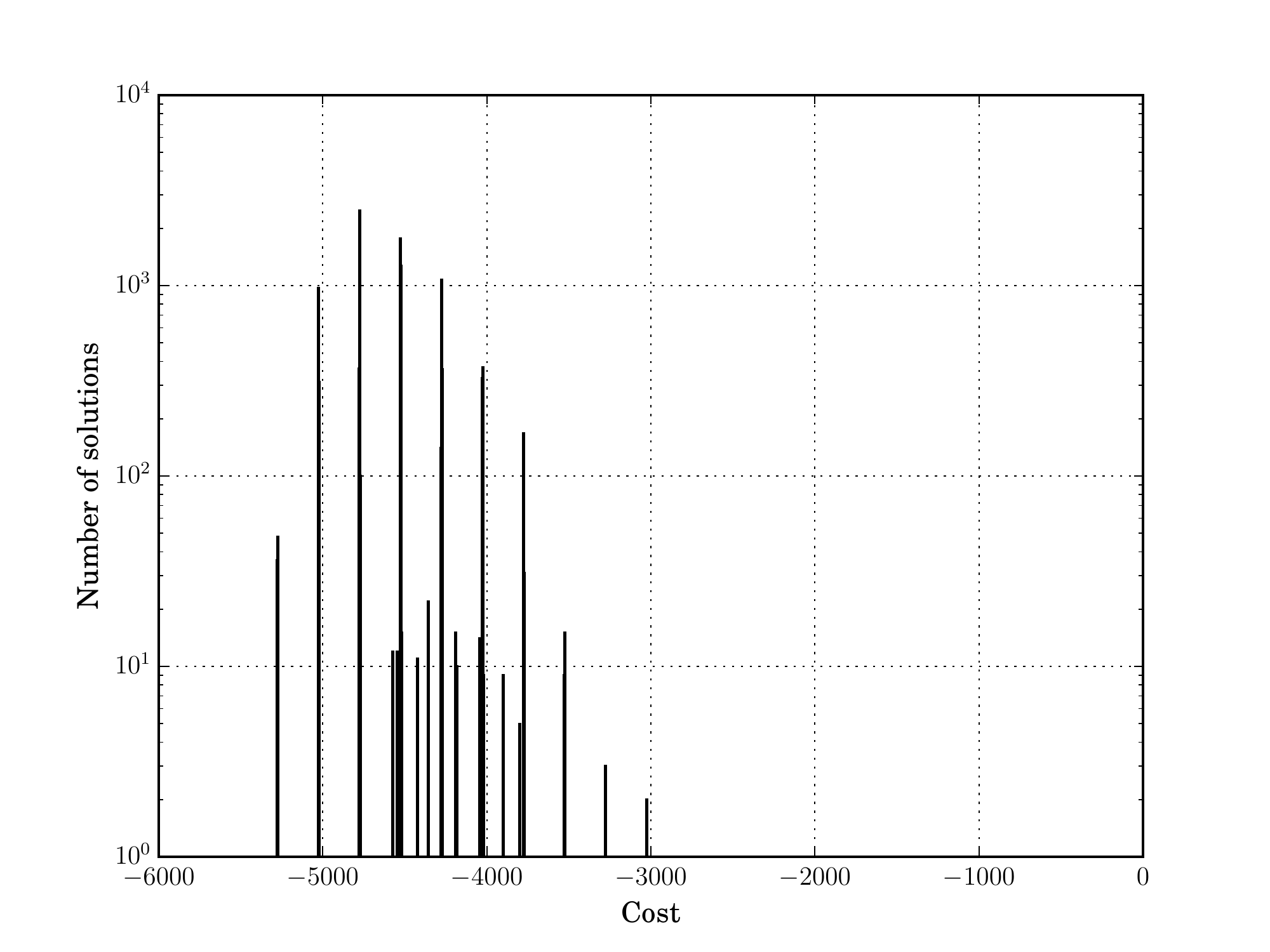}
\end{center}
\caption{Histogram of the economic function over 10000 annealing runs on $G_4$ (with duplication inconsistencies fixed by majority voting).}
\label{fig:G4:correct}
\end{figure}

\paragraph{\textbf{$G_4$.}} This instance leads to a graph with 50 vertices, 125 edges and then to a QUBO with 125 variables and 600 nonzero nondiagonal coefficients. Mapping this QUBO on the D-Wave machine required 951 qubits as shown on Figure \ref{fig:G4:map} (as said previously, this is about $87\%$ of the available qubits for this D-Wave machine). 
Over 10000 runs the optimal solution was never obtained. Still, Figure \ref{fig:G4bw} provides graphic representations of the best, median and worst solutions obtained (respectively with cost $-5524$, $-4526$ and $-2109$). For $G_4$, the optimum value is $-6075$, thus the best solution obtained is around 10\% far off (a better ratio than for $G_3$) and median cost 25\%. Furthermore, neither the best nor the median solution lead to valid matchings since in both, some vertices are covered several times. We also observe that the worst solution has duplication consistency issues. Figure~\ref{fig:G4:histo} shows the histogram of the economic function as outputted by the D-Wave (but renormalized) for the 10000 annealing runs we performed. Additionally, since some of these solutions are inconsistent with respect to duplication, Figure \ref{fig:G4:correct} shows the histogram of the economic function for the solutions in which duplication inconsistencies were fixed by majority voting (thus left shifting the average solution cost from -4492.4 to -4525.8 and the median cost from -4525 to -4526 which is also marginal).

\section{Discussion and perspectives}
\label{sec:discuss}

In this paper, our primary goal was to provide a first study on the behavior of an existing quantum annealer when confronted to old combinatorial beasts known to defeat classical annealing. At the very least, our study demonstrates that these special instances of the maximum (bipartite) matching problem are not at all straightforward to solve on a quantum annealer and, as such, are worth being included in a standard benchmark of problems for these emerging systems. Furthermore, as this latter problem is polynomial (and the specific instances considered in this paper even have straightforward optimal solutions), it allows to precisely quantify the quality of the solutions obtained by the quantum annealer in terms of distance to optimality.

There also are a number of lessons learnt. First, the need for qubit duplication severely limits the size of the problem which can be mapped on the device leading to a ratio between 5 and 10 qubits for 1 problem variable. Yet, a $\approx1000$ qbits D-Wave can tackle combinatorial problems with a few hundred variables, a size which is clearly nontrivial. Also, the need to embed problem constraints (e.g., in our case, matching constraints requiring that each vertex is covered at most once) in the economic function, even with carefully chosen penalty constants, often lead to invalid solutions. This is true both in terms of qubits duplication consistency issues (i.e., qubits representing the same problem variable having different values) as well as for problem specific constraints. This means that operationally using a quantum annealer requires one or more postprocessing steps (e.g., solving qubit duplication inconsistencies by majority voting), including problem-specific ones (e.g., turning invalid matchings to valid ones). 

Of course, the fact that, in our experiments, the D-Wave failed to find optimal solutions for nontrivial instance sizes, does not rule out the existence of an advantage of quantum annealing \emph{as implemented in D-Wave systems} over classical annealing (the existence of which, as previously emphasized, as already been established on specially designed problems \cite{albash-2018}). However, our results tends to rule out (or confirm) the absence of an exponential advantage in the general case of quantum over classical annealing.

Also, since the present study takes a worst-case (instances) point of view, it does not at all imply that D-Wave machines cannot be practically useful, and, indeed, its capacity to anneal in a few tens of $\mu$s makes it inherently very fast compared to software implementations of classical annealing. Stated otherwise, the present study just tends to imply that there are (even non $NP$-hard) problems which are hard for both quantum \emph{and} classical annealing and that on these quantum annealing does not perform significantly better.

In terms of perspectives, it would of course be interesting to test larger instances on D-Wave machines with more qubits. It would also be very interesting to benchmark a device with the next generation of D-Wave qubit interconnection topology (the so-called Pegasus topology \cite{dattani2019pegasus}) which is significantly denser than the Chimera topology. On the more theoretical side of things, trying to port Sasaki and Hajek proof \cite{sasaki-1988} to the framework of quantum annealing, although easier said than done, is also an insighful perspective. Lastly, bipartite matching over the $G_n$ graphs family also gives an interesting playground to study or benchmark emerging classical quantum-inspired algorithms (e.g. Simulated Quantum Annealing \cite{crosson-2016}) or annealers.

\section*{Acknowledgements}

The authors wish to thanks Daniel Est\`eve and Denis Vion, from the Quantronics Group at CEA Paris-Saclay, for their support and fruitul discussions. The authors would also like to warmly thank Pr Daniel Lidar for granting them access to the D-Wave 2X operated at the University of Southern California Center for Quantum Information Science \& Technology on which our experiments were run.


\bibliographystyle{plain}
\bibliography{preprint}

\end{document}